\documentclass[11pt,letterpaper]{article}
\usepackage{booktabs}

\newcommand\spacingset[1]{\renewcommand{\baselinestretch}%
  {#1}\small\normalsize}

\usepackage{latexsym}
\usepackage{amssymb,amsmath, bm,pgfplots,tikz,bbm}
\usepackage{graphicx}
\usepackage{marvosym}
\usepackage{multirow,float}
\usepackage{algpseudocode}
\usepackage{caption}
\usepackage{mathtools}
\usepackage{subcaption}
\usepackage{comment}
\usepackage{enumitem}

\usepackage{natbib}
\usepackage{color}
\usepackage[bookmarksopen=true, bookmarksnumbered=true,
pdfstartview=FitH, breaklinks=true, urlbordercolor={0 1 0}, citebordercolor={0 0 1}]{hyperref}

\usepackage{dcolumn}
\newcolumntype{.}{D{.}{.}{-1}}
\newcolumntype{d}[1]{D{.}{.}{#1}}

\usepackage{theorem}
\theoremstyle{plain}
\theoremheaderfont{\scshape}
\newtheorem{theorem}{Theorem}
\newtheorem{proposition}{Proposition}
\newtheorem{assumption}{Assumption}

\newtheorem{lemma}{Lemma}

\newcommand{\qed}{\hfill \ensuremath{\Box}}
\newcommand{\indep}{\mbox{$\perp\!\!\!\perp$}}

\DeclareMathOperator*{\argmin}{argmin}

\newcommand{\norm}[1]{\left\lVert#1\right\rVert}
\newenvironment{proof}{\vspace{1ex}\noindent{\bf Proof}\hspace{0.5em}}
{\hfill\qed\vspace{1ex}}
\usepackage{kantlipsum}
\allowdisplaybreaks

\graphicspath{{Manuscripts/figs/}}

\usepackage{rotating}

\usepackage{arydshln}
\usepackage{threeparttable}

\usepackage[compact]{titlesec}

\newcommand{\blind}{0}

\usepackage{times}

\usepackage[ruled,linesnumbered,vlined]{algorithm2e}
\usepackage{pifont}

\newtheorem{definition}{Definition}

\newcommand\E{\mathbb{E}}
\newcommand\R{\mathbb{R}}
\newcommand\cR{\mathcal{R}}
\renewcommand\P{\mathbb{P}}
\newcommand\cP{\mathcal{P}}
\newcommand\cQ{\mathcal{Q}}
\newcommand\bq{\bm{q}}

\newcommand\bP{\bm{P}}
\newcommand\bQ{\bm{Q}}

\newcommand\br{\bm{r}}
\newcommand\bR{\bm{R}}

\newcommand\bh{\bm{h}}

\newcommand\bx{\bm{x}}
\newcommand\bX{\bm{X}}

\newcommand\boldf{\bm{f}}
\newcommand\bg{\bm{g}}
\newcommand\bu{\bm{u}}
\newcommand\bU{\bm{U}}
\newcommand\cD{\mathcal{D}}
\newcommand\cT{\mathcal{T}}
\newcommand\cU{\mathcal{U}}

\newcommand\cX{\mathcal{X}}
\newcommand\cY{\mathcal{Y}}
\newcommand\cZ{\mathcal{Z}}
\newcommand\wT{\widetilde{T}}
\newcommand\bgamma{\boldsymbol{\gamma}}
\newcommand\btheta{\boldsymbol{\theta}}
\newcommand\bzeta{\boldsymbol{\zeta}}
\newcommand\blambda{\boldsymbol{\lambda}}
\usetikzlibrary{decorations.markings}
\usetikzlibrary{decorations.pathmorphing}
\usetikzlibrary{shapes.geometric, arrows}
\usetikzlibrary{arrows,decorations.pathmorphing,backgrounds,positioning,fit,matrix}
\usetikzlibrary{shapes,decorations,arrows,calc,arrows.meta,fit,positioning}
\tikzset{auto,node distance =1 cm and 1 cm,semithick,
	state/.style ={circle, draw, minimum width = 0.7 cm},
	point/.style = {circle, draw, inner sep=0.04cm,fill,node contents={}},
	bidirected/.style={Latex-Latex,dashed},
	el/.style = {inner sep=2pt, align=left, sloped}
}

\newcommand{\tit}{Causal Inference with Generative Artificial Intelligence: Application to Texts as Treatments}
\graphicspath{{figs/}} 

\begin{document}

\if\blind1
\title{\tit}
\fi

\if\blind0 

\title{\tit\footnote{An open-source Python package, ``GPI: Generative-AI Powered Inference,'' that implements the proposed methodology is available at \url{https://gpi-pack.github.io/}. The replication code and data are available at \url{https://github.com/k-nakam/gpi_replication}. We thank Christian Fong and Justin Grimmer for kindly sharing their dataset used in Candidate Profile Experiment and Naoki Egami for helpful comments.  We also acknowledge useful comments from Naoki Egami, Max Kasy, and an anonymous reviewer of Harvard's Alexander and Diviya Magaro Peer Pre-Review Program.}}
\author{ Kosuke Imai\thanks{Professor, Department of Government and
      Department of Statistics, Harvard University, Cambridge, MA
      02138. Phone: 617--384--6778, Email:
      \href{mailto:Imai@Harvard.Edu}{Imai@Harvard.Edu}, URL:
      \href{https://imai.fas.harvard.edu}{https://imai.fas.harvard.edu}} \and
  Kentaro Nakamura\thanks{Ph.D. student, John F. Kennedy School of Government, Harvard University. Email: \href{mailto:knakamura@g.harvard.edu}{knakamura@g.harvard.edu}}
}
\fi
\date{\today}
\maketitle\thispagestyle{empty}

\begin{abstract}
  In this paper, we demonstrate how to enhance the validity of causal inference with unstructured high-dimensional treatments like texts, by leveraging the power of generative Artificial Intelligence (GenAI).  Specifically, we propose to use a deep generative model such as large language models (LLMs) to efficiently generate treatments and use their internal representation for subsequent causal effect estimation.  We show that the knowledge of this true internal representation helps disentangle the treatment features of interest, such as specific sentiments and certain topics, from other possibly unknown confounding features.  Unlike existing methods, the proposed GenAI-Powered Inference (GPI) methodology eliminates the need to learn causal representation from the data, and hence produces more accurate and efficient estimates.  We formally establish the conditions required for the nonparametric identification of the average treatment effect, propose an estimation strategy that avoids the violation of the overlap assumption, and derive the asymptotic properties of the proposed estimator through the application of double machine learning.  Finally, using an instrumental variables approach, we extend the proposed GPI methodology to the settings in which the treatment feature is based on human perception.  The GPI is also applicable to text reuse where an LLM is used to regenerate existing texts.  We conduct simulation and empirical studies, using the generated text data from an open-source LLM, Llama~3, to illustrate the advantages of our estimator over state-of-the-art causal representation learning algorithms.

\end{abstract}

\noindent {\bf Key Words:} causal inference, deep generative models, double machine learning, large language models, unstructured treatments

\newpage

\section{Introduction}

The emergence of generative artificial intelligence (GenAI) technology such as large language models (LLMs) has had an enormous impact on many fields, including medicine \citep{LLM_medicine2023}, education \citep{chatGPTeducation2023}, marketing \citep{marketingAI2024}, and social sciences \citep{bisbee_synthetic_2024}.  These tools come with advanced capabilities to generate realistic texts, images, and videos at scale, based on the user-provided prompts.

In this paper, we demonstrate that GenAI can enhance the performance of causal inference based on unstructured data such as text and images. We focus on the problem of estimating the causal effect of a specific treatment feature embedded in text---such as topic or sentiment---while adjusting for other confounding features. Although we assume the treatment feature is pre-specified and measurable, the central challenge lies in learning a low-dimensional representation of the unknown confounders and properly adjusting for them. We show that internal representations extracted from GenAI models can be leveraged within a causal machine learning framework to estimate the causal effects of interest. Specifically, we introduce an experimental design (Section~\ref{sec:example}) and propose the GenAI-Powered Inference (GPI) methodology (Section~\ref{sec:methodology}), which enables efficient estimation of causal effects associated with specific features embedded in unstructured data.

In the proposed experiment, we first generate texts by providing treatment and control prompts to an LLM.  If we wish to use existing texts rather than generate new ones, we ask an LLM to reproduce the same texts exactly.  We then present each generated text to a randomly selected survey respondent and measure their reactions.  Lastly, we directly extract the true internal representation of the generated text from the LLM and analyze it with a machine learning algorithm to disentangle the treatment features from other confounding features contained within the same texts.  The GPI leverages the true vectorized representation of treatment text that is available from an open-source deep generative model.  This enables us to efficiently learn the representation of relevant features in unstructured data even when texts contain strong confounding features.

We establish the nonparametric identification based on the key assumption of {\it separability} between treatment and confounding features.  This assumption states that the treatment feature is not a deterministic function of confounding features and the latter are also not a function of the former.  The assumption is closely related to the concept of disentanglement in the literature \citep{wang_desiderata_2022} and implies that one can intervene the treatment feature without changing confounding features.  We discuss diagnostic tools to detect the potential violation of this assumption.

As part of the proposed estimation approach, we develop a neural network architecture based on TarNet \citep{shalit_estimating_2017} to separately learn treatment and confounding features.  We show that it is possible to nonparametrically identify a {\it deconfounder}, which summarizes all confounding features as a lower-dimensional function of the internal representation obtained from a deep generative model.  Once a deconfounder is estimated, we use it to model the treatment features and estimate the propensity score.  We then apply the double machine learning (DML) methodology to obtain the asymptotically valid confidence intervals \citep{chernozhukov_doubledebiased_2018}. \if \blind0 An open-source Python package, ``GPI: Generative-AI Powered Inference,'' that implements the proposed methodology is available at \url{https://gpi-pack.github.io/}. \fi

To investigate the empirical performance of the proposed methodology, we design simulation studies based on the candidate profile experiment of \cite{fong_discovery_2016} (Section~\ref{sec:simulation}).  In this experiment, survey respondents are asked to evaluate biographies of different political candidates.  Our goal is to infer the causal effects of certain features of these biographies, such as military background and education levels.  We use an open-source LLM, Llama~3, to generate a set of new candidate biographies and design simulation studies to compare the performance of the proposed estimator with that of state-of-the-art methods.  We also have Llama~3 regenerate the existing biographies to examine the performance of the proposed methodology in the case of text reuse.

We find that the GPI outperforms the state-of-the-art methods, which estimate the representation of confounding features using the existing embedding \citep{pryzant_causal_2021, gui_causal_2023}. Specifically, the proposed estimator has a smaller bias and root mean squared error, while its confidence interval retains a proper nominal coverage level.  These findings hold even when the sample size is relatively small, and in the case of text reuse.  Furthermore, we apply the GPI to the candidate profile experiment (Section~\ref{sec:empirical}).  Our analysis shows that the previous military experience of the candidates significantly affects the voter evaluation on average.  Appendix~S6 presents an additional empirical application based on experiments about the public perceptions of US government support for Hong Kong protest \citep{fong_causal_2023}.  We show that the GPI yields much more reasonable estimates than the existing state-of-the-art methods.  \cite{imai2025GPI} presents additional applications of the GPI including one based on images.

Finally, we extend the GPI to the estimation of the causal effects of {\it perceived} treatment features, motivated by the fact that the respondents may perceive the same treatment features differently (Appendix~S3).  A key challenge is that the perceived treatment features may be confounded by possibly unobserved respondent characteristics as well as the confounding features of texts.  To address this, we adopt the instrumental variable approach \citep{imbens_identification_1994}  by using the actual treatment features as instruments for their perceived counterparts.

\paragraph{Related literature.}
Several existing works estimate the causal effects of textual features \citep[see][for a review]{feder_causal_2022}. The key difference between the GPI and these previous approaches is the use of GenAI to produce treatment objects.  We exploit the fact that the {\it true} vectorized representation of generated texts can be obtained directly from open-source LLMs.  In contrast, existing methods must learn representation from the treatment texts.  For example, \cite{fong_discovery_2016} and \cite{ahrens_bayesian_2021} impose topic models, whereas \cite{pryzant_causal_2021} and \cite{gui_causal_2023} estimate the vector representation using the BERT (Bidirectional Encoder Representations from Transformers) embeddings.  We show that the use of true representation not only improves the estimation performance but also significantly increases computational efficiency.

We advance the literature on causal inference with text by formalizing the assumptions necessary for causal identification. Most prior studies implicitly assume that confounding features are not functions of the treatment feature \citep[e.g.,][]{pryzant_causal_2021, gui_causal_2023}, with the exception of \citet{daoud_conceptualizing_2022}, who states this assumption explicitly. Consequently, no existing estimation method directly incorporates this assumption. For instance, \citet{fong_causal_2023} recommends using topic models and adjusting only for those estimated topics that are not functions of the treatment feature. In practice, however, reliably identifying such topics is challenging, since any topic may be entangled with the treatment feature. Relatedly, the representation learning literature discusses this condition under the label of disentanglement \citep{wang_desiderata_2022}, but does not consider interventions on the treatment feature while holding confounding features fixed. To address this gap, we introduce the separability assumption, which implies that it be possible to intervene on the treatment feature without altering the confounding features.

Our work has broader implications for the literature on causal inference and unstructured objects in general.  For example, scholars have developed methods that adjust for texts as confounders, but all of these methods estimate a low-dimensional representation of confounding information from texts \citep{veitch_adapting_2020, roberts_adjusting_2020, klaassen_doublemldeep_2024, mozer_matching_2020, mozer_leveraging_2024}. Although our paper focuses on the use of texts as treatments rather than confounders, the proposed use of GenAI should be beneficial for these other cases where existing estimation methods are likely to suffer from estimation error \citep{keith_text_2020}.
Similarly, the GPI can also be applied to causal inference problems with images and even videos. For example, \cite{jerzak_integrating_2023, jerzak_image-based_2023} have considered the use of causal inference with images in an observational setting. Given the availability of deep generative models for images, it would be of interest to use GenAI for improved causal inference with images \citep[e.g.,][]{
  ramesh_zero-shot_2021, rombach_high-resolution_2022}.

Our work also contributes to the literature on representation learning. Since unstructured objects such as texts are high-dimensional, learning a low-dimensional representation is crucial \citep[e.g.,][]{shi_adapting_2019, wang_desiderata_2022}. Some propose learning a low-dimensional representation that predicts both treatment and outcome \citep[e.g.,][]{veitch_adapting_2020, gui_causal_2023}. Although we also use a low-dimensional representation to adjust for confounding features, the GPI disentangles confounding features from treatment features without violating the overlap assumption.  In addition, the GPI focuses on estimating causal effects rather than discovering of causal structure, which is an important goal of representation learning \citep{Scholkopf2021TowardCausal}.

Finally, the GPI differs from that of \cite{wang:blei:19} though both use the estimated ``deconfounder'' for confounding adjustment.  \citeauthor{wang:blei:19} adjust the {\it unobserved} confounding by estimating the deconfounder as a function of treatments \citep{imai:jian:19}.  In contrast, the GPI learns a representation of {\it observed} confounding by estimating the deconfounder as a function of generative model's internal representation of treatment objects.  In this setting, we formally establish the nonparametric identification of the average treatment effect.

\section{The Experimental Design for Texts as Treatments} \label{sec:example}

While the G{PI is a general methodological approach, it is helpful to consider a concrete application.  We use the candidate profile experiment of \cite{fong_discovery_2016} which investigates how various features of a political candidate affect voter evaluation.  In the context of this experiment, we describe our alternative approach that uses LLMs to generate treatment texts.  In Sections~\ref{sec:simulation}~and~\ref{sec:empirical}, we return to this application to empirically evaluate the performance of the GPI.

\subsection{Candidate profile experiment}

\cite{fong_discovery_2016} collected 1,246 candidate biographies (written in texts) from Wikipedia, and then asked a total of 1,886 voters to answer an online survey evaluating up to four randomly assigned candidate profiles.  Specifically, the survey asked these voters to rate each candidate biography using a feeling thermometer that ranges from 0 (cold) to 100 (warm).  The authors
first used a supervised Bayesian model based on the Indian buffet process to discover a total of 10 treatment features and then estimated the marginal association between each treatment feature and the observed feeling thermometer value.

Unlike the original analysis, we consider a setting in which researchers have a specific treatment feature of interest.  Suppose that we are interested in estimating the causal effect of having a military background on voter evaluation.  The relevant political science literature suggests that the experience and occupation of candidates play an important role \citep[e.g.,][]{kirkland_candidate_2018, campbell_what_2014, pedersen_voter_2019}.  Indeed, the original analysis shows that candidate biographies with military background tend to receive a high feeling thermometer score.

The challenge, however, is that military background may be correlated with other features present in candidate biographies.   Table~S1 of Appendix~S1 displays two example biographies used in the original experiment.  The first describes a candidate with military background, whereas the second shows another without military background.  Yet, these two biographies also differ in terms of other features, including educational background, marital status, and family structure.  The length of each biography and their levels of detail are also different.  If these differences are correlated with military background and influence voter evaluation, a simple comparison between biographies with military background and those without it would lead to biased causal estimates.

\subsection{Using large language models to (re)generate treatment texts}
\label{subsec:reuse}

We use an LLM to generate treatment texts.  In the current context, we can achieve this in two ways.  First, we can provide a prompt to an LLM, asking it to generate a candidate biography from scratch.  Alternatively, we can use existing biographies (e.g., those collected by \cite{fong_discovery_2016}) and then ask an LLM to reproduce the same biographies without any modification.

Both approaches require (1) the coding of the treatment variable (e.g., the existence of military background) and (2) the extraction of the internal representation of LLM used to generate treatment texts.  The first requirement may mean that human coders have to read treatment texts unless one is willing to assume that LLM has a good compliance with the instruction in the prompts.  The second requirement implies that we should use open-source LLMs including GPT (Generative Pre-trained Transformer), 
Llama (Large Language Model Meta AI), 
and OPT (Open Pre-trained Transformer). 

Table~S2 of Appendix~S1 shows an example from each approach.  Here, we use Meta's Llama~3 instruction-tuned model with eight billion parameters.  This model takes two types of inputs: system-level inputs ({\bf System}), which define the type of task to be performed, and user-level inputs ({\bf User}), which define a specific task to be performed.  The first example in the table shows how the model generates a new candidate biography with military background from scratch, whereas the second shows how the model reproduces a given biography.  The former suggests that an LLM can create realistic treatments, while the latter indicates that it can follow the instruction accurately.  

We emphasize that the use of LLM in itself does not automatically solve the confounding bias. This is because the LLM learns the associations of words in real-world texts, and even if one manipulates the concepts with instructions, other correlated concepts might also be influenced, causing the confounding bias \citep{hu_causal_2021}. 

\section{The Proposed Methodology}
\label{sec:methodology}

We turn to the proposed GPI methodology that adjusts for confounding features in unstructured treatment objects such as texts to estimate the causal effects of the specific treatment feature. We begin by defining the causal quantity of interest, establish its nonparametric identification, and then develop an estimation strategy.

\subsection{Assumptions and causal quantity of interest}

Consider a simple random sample of $N$ respondents drawn from a population of interest.  For each respondent $i=1,2,\ldots,N$, we assign a prompt $\bP_i$ that is randomly and independently sampled from a set of potential prompts $\cP$.  In our application, the prompts are based on natural language (e.g., ``Create a biography of an American politician who has some military experience'').  
Given each prompt, we use a deep generative model to generate a {\it treatment object} such as text, denoted by $\bX_i \in \cX$ where $\cX$ is the support of $\bX_i$. 

We use a broad definition of a deep generative model to encompass LLMs and other foundation models.  Indeed, our definition includes many models for texts \citep[e.g.,][]{touvron_llama_2023, zhang_opt_2022, jiang_mistral_2023} and images \citep[e.g.,][]{ramesh_zero-shot_2021, rombach_high-resolution_2022}.

\begin{definition}[Deep Generative Model]\label{deep_use} \spacingset{1} A deep generative model is the following probabilistic model that takes prompt $\bP_i$ as an input and generates the treatment object $\bX_i$ as an output:
$$
\begin{aligned}
&\P(\bX_i  \mid \bh_{\bgamma}(\bR_i))\\
&\P(\bR_i \mid \bP_i)
\end{aligned}
$$
where $\bR_i \in \cR \subset \R^{d_R}$ denotes an observable internal representation of $\bX_i$ contained in the model and $\bh_{\bgamma}(\bR_i)$ is a deterministic function parameterized by $\bgamma$ that completely characterizes the conditional distribution of $\bX_i$ given $\bR_i$.  
\end{definition}
Under this definition, $\bR$ represents a lower-dimensional representation of the treatment object $\bX$ and is a hidden representation of neural networks.  In addition, $\bh_{\bgamma}(\bR)$ is the propensity function \citep{imai_causal_2004} and is both known and observable in an open-source deep generative model.  Thus, the treatment object $\bX$ depends on $\bP$ only through $\bh_{\bgamma}(\bR)$. Note that any given text $\bX_i$ can have multiple internal representations.  For example, one can first instruct an LLM to generate a new text and then ask it to repeat the generated text exactly.  These two prompts will produce the same text but yield different internal representations.  In Section~\ref{subsec:sim-results}, our simulation study shows that the internal representation of regenerated text leads to more efficient causal estimates. 

In the proposed experiment, each respondent $i$ is exposed to the generated treatment object $\bX_i$, and subsequently generates the outcome variable $Y_i \in\cY \subset \mathbb{R}$ where $\cY$ is its support.  In our application, the treatment object is a candidate biography and the outcome is a respondent's evaluation of the biography.  Let $Y_i(\bx)$ denote the potential outcome of respondent $i$ when exposed to a treatment object $\bx \in \cX$.   Then, the observed outcome is solely determined by the assigned treatment object, i.e., $Y_i = Y_i(\bX_i)$.  
This experimental design implies the following two assumptions. 
\begin{assumption}[Consistency] \label{consistency} \spacingset{1} The potential
    outcome under the treatment object $\bx \in \cX$, denoted by
    $Y_i(\bx)$,  equals the observed outcome $Y_i$ under the
    realized treatment object $\bX_i$:
    $$Y_i \ = \ Y_i(\bX_i).$$
\end{assumption}
  
\begin{assumption}[Random Assignment of Prompt] \label{ignorability} \spacingset{1}
  Prompt is randomly assigned such that the following independence
  holds for all $i$ and $\bx \in \cX$:
  $$Y_i(\bx) \ \indep \ \bP_i.$$
\end{assumption}

We estimate the causal effect of a particular feature that can be part of a treatment object.  For simplicity, we consider a binary treatment feature denoted by $T_i$ for each $i$.  In our application, $T_i = 1$ if candidate biography $i$ contains military background and $T_i = 0$ otherwise.  We assume that the treatment feature is determined solely by the treatment object \citep{egami_how_2022}.
\begin{assumption}[Treatment Feature]\label{det_trt} \spacingset{1}
There exists a deterministic function $g_{T}: \cX \to \{0,1\}$ that maps a treatment object $\bX_i$ to a binary treatment feature of interest $T_i$, i.e., $$T_i \ = \ g_T(\bX_i).$$
\end{assumption}
This assumption is violated, for example, if respondents infer different values of the treatment feature from the same treatment object.   We address this issue in Appendix~S3 by considering perceived treatment, which reflects heterogeneity between respondents.

Next, we define confounding features, which represent all features of $\bX$ other than the treatment feature $T$ that influence the outcome $Y$.  These confounding features, denoted by $\bU \in \cU$, are based on a vector-valued deterministic function of $\bX$ where $\cU$ denotes their support.  The confounding features may be multidimensional, though we assume that its dimensionality is much smaller than that of the treatment object. In our application, confounding features may include other candidate characteristics and textual features of biographies, such as their length.   These confounding features are possibly correlated with the treatment feature.  Unlike the treatment feature, however, we do not directly observe the confounding features.  Formally, the confounding features are defined as follows \citep{fong_causal_2023, pryzant_causal_2021}.
\begin{assumption}[Confounding Features]\label{confounder} \spacingset{1}
  There exists an unknown vector-valued deterministic function $\bg_{\bU}: \cX \to \cU$ that maps an unstructured object $\bX_i \in \cX$ to the confounding features $\bU_i \in \cU$, i.e.,
  $$\bU_i \ = \ \bg_{\bU}(\bX_i),$$
  where $\dim(\bU_i) \ll \dim(\bX_i)$.
\end{assumption}

Finally, we introduce our key assumption that the potential outcome is a function of treatment and confounding features and that we can intervene the treatment feature without changing the confounding features.  In other words, the treatment feature cannot be represented as a deterministic function of the confounding features. In addition, confounding features should not vary as a deterministic function of treatment feature in order to avoid ``posttreatment'' bias \citep{daoud_conceptualizing_2022}. In the candidate biography application, the assumption essentially implies that researchers need to be able to imagine hypothetical candidate biographies with and without military background while keeping the confounding features constant.  We formalize this assumption as follows.
\begin{assumption} {\sc (Separability of Treatment and Confounding Features)} \label{separability} \spacingset{1} The potential outcome is a function of the treatment feature of interest $t$ and another separate function of the confounding features $\bu$.  That is, for any given $\bx \in \cX$ and all $i$, we have:
$$
Y_i(\bm{x}) = Y_i(t, \bu)=Y_i(g_{T}(\bx), \bg_{\bU}(\bx)),
$$
where $t=g_{T}(\bx) \in \{0,1\}$ and $\bu=\bg_{\bU}(\bx) \in \cU$. In addition, $g_T$ and $\bg_{\bU}$ are separable.  That is, there exists no deterministic function $\tilde{g}_T: \cU \to \{0,1\}$, which satisfies $g_T(\bx) = \tilde{g}_T (\bg_{\bU}(\bx))$ for some $\bx \in \cX$.  Similarly, there exist no deterministic functions $\bg^\prime: \cX \to \cX^\prime$ and $\tilde{\bg}_{\bU}: \{0,1\} \times \cX^\prime \to \cU$, which satisfy $\bg_{\bU}(\bx) = \tilde{\bg}_{\bU} ( g_T(\bx), \bg^\prime(\bx))$ for all $\bx \in \cX$ and $\tilde{\bg}_{\bU} ( 1, \bg^\prime(\bx^\prime))\ne \tilde{\bg}_{\bU} (0, \bg^\prime(\bx^\prime))$ for some $\bx^\prime \in \cX$.  \end{assumption}

The first requirement of separability (i.e., no existence of $\tilde{g}_T$) implies that the treatment should not be a deterministic function of confounding features, whereas the second requirement (i.e., no existence of $\bg^\prime$ and $\tilde\bg_{\bU}$) means that the confounding features cannot vary as a deterministic function of treatment.  Thus, the separability assumption holds when the treatment feature does not completely overlap with the other features that influence the outcome (i.e., confounding features). In Appendix~S2, we provides two simple examples: one, in which the separability assumption holds, and the other one, where it is violated.

As shown in Section~\ref{subsec:identification}, Assumption~\ref{separability} plays an essential role in the identification of causal effects.  
Importantly, Assumptions~\ref{det_trt}--\ref{separability} imply the standard overlap condition in causal inference, which enables to identify causal effects without extrapolation.  In practice, as demonstrated in Section~\ref{subsec:sim-results}, the violation of Assumption~\ref{separability} can be diagnosed by examining if the estimated propensity scores take extreme values.  The following lemma formally establishes that the separability assumption implies the overlap condition.
\begin{lemma}[Overlap] \label{overlap} \spacingset{1} Under Assumptions~\ref{det_trt}--\ref{separability}, for any $t \in \{0,1\}$ and $\bu \in \cU$, $$\P(T_i = t \mid \bU_i = \bu) > 0. $$
\end{lemma}
The proof is given in Appendix~S4.1.  

Under this setup, we estimate the average causal effect of the treatment feature while controlling for the confounding features.   This average treatment effect (ATE) is defined as follows:
\begin{equation}
      \tau \ :=  \ \E[Y_i(1, \bU_i) - Y_i(0, \bU_i)]. \label{eq:ATE}
\end{equation}

\subsection{Nonparametric identification}
\label{subsec:identification}

\begin{figure}[t]
\centering \spacingset{1}
\begin{tikzpicture}
\node[text centered] (l) {$\bP$};
\node[right = 1.5 of l, text centered] (r) {$\bR$};
\node[right = 1 of r, text centered] (h) {$\bh_{\bgamma}(\bR)$}; 
\node[right = 1.5 of h, text centered] (x) {$\bm{X}$};
\node[below right = 1 of x, text centered] (u) {$\bU = g_{\bU}(\bX)$};
\node[above right = 1 of x, text centered] (t) {$T = g_{T}(\bX)$};
\node[right= 4 of x, text centered] (y) {$Y$};
\node[rectangle, draw, minimum width = 4.25cm, minimum height = 1.5cm] (z) at (3.25,0) {}; 
\node[rectangle, minimum width = 4.25cm, minimum height = 1.5cm] (z) at (3.25,1) {Deep generative model}; 
\draw[-Stealth] (l) --  (r);
\draw[-Stealth, double, color = red] (r) --  (h); 
\draw[-Stealth, double, color = red] (h) --  (x); 
\draw[-Stealth, double, color = red] (x) --  (u);
\draw[-Stealth, double, color = red] (x) --  (t);
\draw[-Stealth] (u) --  (y);
\draw[-Stealth] (t) --  (y);
\end{tikzpicture}
\caption{Directed Acyclic Graph of the Assumed Data Generating Process. A treatment object $\bX$ is generated using a deep generative model (rectangle), in which a prompt $\bP$ produces an internal representation $\bR$ that generates $\bX$ through a propensity function  $\bh_{\bgamma}(\bR)$. The treatment object affects the outcome $Y$ through its treatment feature of interest $T$ and other confounding features $\bU$. An arrow with red double lines represents a deterministic causal relation while an arrow with a single line indicates a possibly stochastic relationship.}
\label{DAG}
\end{figure}
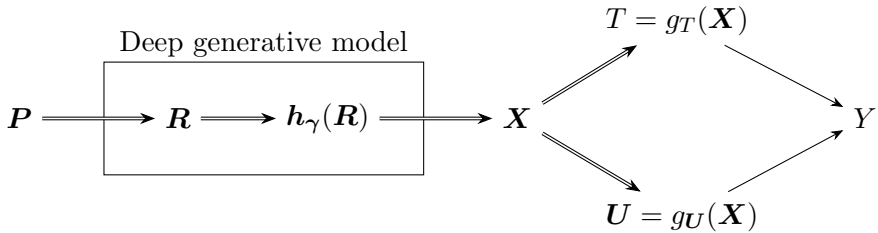

Figure~\ref{DAG} presents a directed acyclic graph (DAG) that summarizes the data generating process described above.  In the DAG, an arrow with double lines represents a deterministic causal relationship, while an arrow with a single line indicates a possibly stochastic causal relationship.  We consider a deep generative model whose decoding is deterministic, meaning that the output object is a deterministic function of the input prompt. Formally, we make the following assumption.
\begin{assumption}[Deterministic Decoding]\label{det_dec} \spacingset{1} The output layer of a deep generative model is deterministic.  That is, $\P(\bX_i \mid \bh_{\bgamma}(\bR_i))$ in Definition~\ref{deep_use} is a degenerate distribution.
\end{assumption}

If $\P(\bX_i \mid \bh_{\bgamma}(\bR_i))$ is stochastic, the noise introduced by the deep generative model can confound the treatment in an unknown way.  Fortunately, almost all LLMs have the option of deterministic decoding (e.g., greedy, beam, and contrastive searches), thereby easily satisfying  the assumption \citep[e.g.,][]{su_contrastive_2022}. For example, for greedy decoding, we typically only need to set the temperature parameter to zero, instructing a model to always produce the same texts.  Indeed, many LLMs also have deterministic {\it encoding} architectures, making the entire text generation process deterministic. Similarly, for images, we can make the final decoding step deterministic for stochastic diffusion models, including Stable Diffusion \citep*{rombach_high-resolution_2022}.  

Given this setup, we establish the nonparametric identification of the marginal distribution of potential outcome. We prove the existence of a deconfounder function $\boldf(\bR_i)$ that satisfies the conditional independence relation $Y_i \indep \bR_i \mid T_i, \boldf(\bR_i)$.  One trivial example of the deconfounder is the confounding features $\bU_i$, which are a deterministic function of $\bR_i$ under Assumption~\ref{det_dec}. But, the deconfounder need not be unique.  In fact, we show that it is possible to identify the marginal distribution of potential outcome by adjusting for any deconfounder. 
\begin{theorem} {\sc (Nonparametric Identification of the Marginal Distribution of Potential Outcome)}\label{iden_det} \spacingset{1}
  Under Assumptions~\ref{consistency}--\ref{det_dec}, there exists a deconfounder function $\boldf: \cR \rightarrow \cQ \subset \R^{d_Q}$ with $d_Q = \dim(\cQ) \le d_R=\dim(\cR)$ that satisfies the following conditional independence relation: \begin{equation}
  Y_i \indep \bR_i \mid T_i = t, \boldf(\bR_i) = \bq, \label{eq:YindepRgivenTf}
\end{equation}
where $0<\P(T_i = t \mid \boldf(\bR_i)=\bq)<1$ for all $t=0,1$ and $\bq \in \cQ$. In addition, the treatment feature and a deconfounder are separable.  By adjusting for such a deconfounder, we can uniquely and nonparametrically identify the marginal distribution of the potential outcome under the treatment condition $T_i = t$ for $t\in \{0,1\}$ as:
\begin{align*}
  \P(Y_i(t, \bU_i) = y) \ = \ \int_{\cR} \P(Y_i = y \mid T_i = t, 
  \boldf(\bR_i)) dF(\bR_i),
\end{align*}
for all $t\in \{0,1\}$ and $y \in \cY$. 
\end{theorem}
The proof is given in Appendix~S4.2.  The definition of separability is given in Assumption~\ref{separability} for the treatment and confounding features given the treatment object.  In Theorem~\ref{iden_det}, we apply the same definition to the treatment feature and a deconfounder given the internal representation.  Specifically, let $f_T: \cR \rightarrow \{0,1\}$ be the function that maps the internal representation to the treatment feature.  Such a function exists because the treatment feature is a deterministic function of treatment object, which is in turn a deterministic function of the internal representation by Assumption~\ref{det_dec}.  Then, we assume that $f_T$ and $\boldf$ are separable.  That is, there exists no deterministic function $\tilde{f}_T: \cQ \to \{0,1\}$, which satisfies $f_T(\br) = \tilde{f}_T (\boldf(\br))$ for all $\br \in \cR$.  Similarly, there exist no deterministic functions $\boldf^\prime: \cQ \to \cQ^\prime$ and $\tilde{\boldf}: \{0,1\} \times \cQ^\prime \to \cQ$, which satisfy $\boldf(\br) = \tilde{\boldf} ( f_T(\br), \boldf^\prime(\br))$ for all $\br \in \cR$ and $\tilde{\boldf} ( 1, \boldf^\prime(\br^\prime))\ne \tilde{\boldf} (0, \boldf^\prime(\br^\prime))$ for some $\br^\prime \in \cR$. The identification of the ATE defined in Equation~\eqref{eq:ATE} can be established in the same way, except that the deconfounder assumption is relaxed to the mean independence condition
\begin{align}
  \E[Y_i \mid T_i = t, \boldf(\bR_i)] = \E[Y_i \mid T_i = t, \bR_i] \label{eq:YindepRgivenTfmean},
\end{align}
which we will use for estimation in the next section.

We emphasize that one cannot directly adjust for the internal representation of the treatment object.  Such direct adjustment leads to the lack of overlap because, under Assumptions~\ref{det_trt}~and~\ref{det_dec}, the treatment feature $T_i$ is a deterministic function of $\bR_i$.  In addition, since the deconfounder $\boldf(\bR_i)$ is typically of much lower dimension than the internal representation of the treatment object $\bR_i$, making adjustments for the former leads to a more effective estimation strategy.

\subsection{Estimation and inference}
\label{subsec:estimation}

Given the identification result, we next consider estimation and inference.  Our estimation strategy is based on the following two observations.  First, Assumption~$\ref{separability}$ implies that the deconfounder $\boldf(\bR_i)$ should not be a function of the treatment feature $T_i$.  Second, the deconfounder should satisfy the conditional mean independence relation given in Equation~\eqref{eq:YindepRgivenTfmean}.  For simplicity, we assume independence across observations.  Technically, if a deep generative model has a stochastic component, we can only assume the conditional mean independence of observations given the internal representation.  However, as discussed earlier, many language models have the option of making the whole text generation process deterministic, guaranteeing this conditional mean independence.

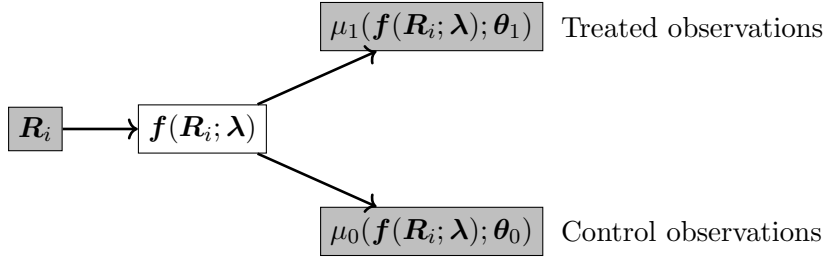
\begin{figure}[t]
\centering \spacingset{1}
    \begin{tikzpicture}[
    node distance=0.5cm and 0.5cm,
    box/.style={draw, rectangle, minimum height=0.5cm, minimum width=0.5cm},
    func/.style={draw, rectangle, minimum size=0.5cm},
    ]
    
    \node[box] (z1) {$\bR_i$};
    \node[box, right = 1.0 of z1, fill = lightgray] (z3) {$\boldf(\bR_i; \blambda)$};
    \node[box, above right= 1.0 of z3, fill = lightgray] (t13) {$\mu_1(\boldf(\bR_i; \blambda); \btheta_1)$};
    \node[box, below right= 1.0 of z3, fill = lightgray] (t03) {$\mu_0(\boldf(\bR_i; \blambda); \btheta_0)$};

    \node[box, right=1.0 of t13] (t1label) {$Y_i\mid T_i = 1$: Treated observations};
    \node[box, right=1.0 of t03] (t0label) {$Y_i \mid T_i = 0$: Control observations};
    
    \draw[->, line width= 1] (z1) --  (z3);
    \draw[->, line width= 1] (z3) --  (t03);
    \draw[->, line width= 1] (z3) --  (t13); 
    \draw[->, line width= 1] (t13) --  (t1label); 
    \draw[->, line width= 1] (t03) --  (t0label); 
\end{tikzpicture}
\caption{Diagram Illustrating the Proposed Model Architecture. The proposed model takes an internal representation of a treatment object $\bR_i$ as an input, and finds a deconfounder $\boldf(\bR_i)$, which is a lower-dimensional representation of $\bR_i$, and then use it to predict the conditional expectation of the outcome $\mu_t(\boldf(\bR_i)) := \E[Y_i \mid T_i = t,\boldf(\bR_i)]$ under each treatment arm $t=0,1$.}
\label{tarnet}
\end{figure}

We use a neural network architecture based on TarNet \citep{shalit_estimating_2017} to estimate the conditional potential outcome function given the deconfounder, i.e.,
$$\mu_t(\boldf(\bR_i)) \ := \ \E[Y_i(t, \bU_i) \mid \boldf(\bR_i)], \quad \text{ for } t=0,1.$$
Our architecture, which is summarized in Figure~\ref{tarnet}, simultaneously estimates the deconfounder and the outcome model.  Specifically, we minimize the following loss function:
\begin{equation}
  \{\hat\blambda, \hat\btheta_0, \hat\btheta_1\} \ = \ \argmin_{\blambda, \btheta_0, \btheta_1}  \frac{1}{n}\sum_{i = 1}^n \left\{Y_i - \mu_{T_i}(\boldf(\bR_i; \blambda); \btheta_{T_i})\right\}^2, \label{loss_text_original}
\end{equation}
where $n$ is the sample size.  We make the parameters of neural network explicit by letting $\blambda$ represent the parameters of deconfounder $\boldf$ to be estimated, and using $\btheta_{t}$ to denote the parameters of the nuisance function $\mu_t$. 

Given the above architecture, we estimate the ATE using the double machine learning (DML) framework of \cite{chernozhukov_doubledebiased_2018}, in which both the outcome and the propensity score models are estimated. Here, we estimate the propensity score model as a function of the estimated deconfounder, i.e., $\pi(\boldf(\bR_i; \hat\blambda)) = \Pr(T_i = 1 \mid \boldf(\bR_i; \hat\blambda))$ after solving the minimization problem in Equation~\eqref{loss_text_original}.  Crutially, we do not model the propensity score as a function of internal representation $\bR_i$ to avoid violating the assumption of overlap.  Thus, the deconfounder only captures the features of treatment object that are predictive of the outcome beyond the treatment, including the features that are completely unrelated to the treatment.  The deconfounder, however, will not capture the features of treatment object that affect the outcome only through the treatment feature.  So long as such features exist (Assumption~\ref{separability}), the overlap assumption will hold (Lemma~\ref{overlap}).

In the representation learning literature, DragonNet \citep{shi_adapting_2019} is a popular estimation method used by many researchers \citep[see e.g.,][as well as \citealt{pryzant_causal_2021} who also use it in their implementation code]{veitch_adapting_2020, gui_causal_2023}.  Unlike our approach, DragonNet includes the cross-entropy loss between the propensity score model and the treatment variable when estimating the deconfounder $\boldf(\bR_i)$.  However, this joint estimation leads to $\P(T_i = 1 \mid \boldf(\bR_i)) = \P(T_i = 1 \mid \bR_i)$ due to the fact that $\boldf(\bR_i)$ is a balancing score satisfying $T_i \indep \bR_i \mid \boldf(\bR_i)$.  This is problematic because $\P(T_i = 1 \mid \bR_i)$ is degenerate under Assumptions~\ref{det_trt}~and~\ref{det_dec}.  Thus, we first estimate the deconfounder using Equation~\eqref{loss_text_original} and then model the propensity score given the estimated deconfounder.

In sum, the entire estimation procedure can be described as follows. Denote the observed data by $\cD:=\{\cD_i\}_{i=1}^N$ where $\cD_i := \{Y_i, T_i, \bR_i\}$.  We use the following $K$-fold cross-fitting procedure, assuming that $N$ is divisible by $K$.
\begin{enumerate} \spacingset{1}
    \item Randomly partition the data into $K$ folds of equal size where the size of each fold is $n = N/K$.  The observation index is denoted by $I(i)\in \{1,\dots,K\}$ where $I(i)=k$ implies that the $i$th observation belongs to the $k$th fold.
    \item For each fold $k \in \{1, \cdots, K\}$, use observations with $I(i)\ne k$ as training data:
    \begin{enumerate} \sloppy
    \item split the training data into two folds, $I_1^{(-k)}$ and $I_2^{(-k)}$
    \item simultaneously obtain an estimated deconfounder and an estimated conditional outcome function on the first fold, which are denoted by $\hat\boldf^{(-k)}(\{\bR_i\}_{i \in I_1^{(-k)}}):=\boldf(\{\bR_i\}_{i \in I_1^{(-k)}}; \hat\blambda^{(-k)})$ and $\hat{\mu}_t^{(-k)}(\{\bR_i\}_{i \in I_1^{(-k)}}):=\mu_t(\boldf(\{\bR_i\}_{i \in I_1^{(-k)}}; \hat\blambda^{(-k)}); \hat\btheta^{(-k)})$, respectively, by solving the optimization problem given in Equation~\eqref{loss_text_original}, and
    \item obtain an estimated propensity score given the estimated deconfounder on the second fold, which is denoted by $\hat{\pi}^{(-k)}(\hat\boldf^{(-k)}(\{\bR_i\}_{i \in I_2^{(-k)}})):=\hat{\pi}^{(-k)} (\boldf(\{\bR_i\}_{i \in I_2^{(-k)}}; \hat\blambda^{(-k)}))$.
    \end{enumerate}
  \item Compute the GPI estimator $\hat{\tau}$ as a solution to:
    \begin{equation}
      \frac{1}{nK}\sum_{k = 1}^K  \sum_{i: I(i) =k} \psi(\cD_i; \hat{\tau}, \hat{\boldf}^{(-k)}, \mu^{(-k)}_1, \mu^{(-k)}_0, \hat\pi^{(-k)}) \ = \ 0, \label{eq:tau.hat}
      \end{equation}
      where
      \begin{equation}
        \begin{aligned}
        & \psi(\cD_i; \tau, \boldf, \mu_1, \mu_0, \pi)\\
        & \ = \  \frac{T_i \{Y_i - \mu_1(\boldf(\bR_i))\}}{\pi(\boldf(\bR_i))} - \frac{( 1- T_i) \{Y_i - \mu_0(\boldf(\bR_i))\}}{1 - \pi(\boldf(\bR_i))}  + \mu_1(\boldf(\bR_i)) - \mu_0(\boldf(\bR_i)) - \tau. 
        \end{aligned} \label{eq:psi}
        \end{equation}
\end{enumerate}

To derive the asymptotic properties of the proposed GPI estimator, we assume the following regularity conditions.

\begin{assumption}[Regularity Conditions]\label{reg_text} \spacingset{1} Let $c_1$, $c_2$, and $q > 2$ be positive constants and $\delta_n$ be a sequence of positive constants approaching zero as the sample size $n$ increases.  Then, the following conditions hold.
  \begin{enumerate}[label=(\alph*)]
    \item \label{reg_text_primitive} (Primitive conditions) 
    \begin{align*}
        &\E[|Y_i|^q]^{1/q} \leq c_1, \ \ \sup_{\br \in \mathcal{R}}\E [|Y_i - \mu_{T_i}(\boldf(\br) )|^2 \mid \bR_i = \br] \leq c_1, \ \ \E[ |Y_i - \mu_{T_i}(\boldf(\bR_i))|^{2} ]^{1/2} \geq c_2.
    \end{align*}
  \item (Outcome model estimation) \label{reg_text_outcome}
    \begin{align*}
        &\E[|\hat{\mu}_{T_i}(\hat{\boldf}(\bR_i)) - \mu_{T_i}(\boldf(\bR_i))|^q]^{1/q} \leq c_1, \quad \E[|\hat{\mu}_{T_i}(\hat{\boldf}(\bR_i)) - \mu_{T_i}(\boldf(\bR_i))|^2]^{1/2} \leq \delta_n n^{-1/4}.
    \end{align*}
  \item (Deconfounder estimation) \label{reg_text_deconfounder}
   \begin{align*}
      & \E\left[\norm{\hat{\boldf}(\bR_i) - \boldf(\bR_i)}^q\right]^{1/q} \leq c_1, \quad \mathbb{E}\left[ \norm{\hat{\boldf}(\bR_i) - \boldf(\bR_i)}^2 \right]^{1/2} \leq \delta_n n^{-1/4}
    \end{align*}    
  \item (Propensity score estimation) \label{reg_text_propensity} $\pi(\cdot)$ is Lipschitz continuous at the every point of its support, and satisfies:
    \begin{align*}
        & \E[|\hat{\pi}(\boldf(\bR_i)) - \pi(\boldf(\bR_i))|^q]^{1/q} \leq c_1, \quad  \E[|\hat{\pi}(\boldf(\bR_i)) - \pi(\boldf(\bR_i))|^2]^{1/2} \leq \delta_n n^{-1/4}.
    \end{align*}
  \end{enumerate}
\end{assumption}
Like the standard application of DML, the required rate for nonparametric estimation of nuisance models is slower than the usual parametric estimation rate of $n^{-1/2}$.  Recall that we use neural networks for the joint estimation of outcome model and deconfounder.  The required convergence rate of $n^{-1/4}$ is achievable with the standard neural network architecture \citep*{farrell_deep_2021}. The propensity score function can be estimated using various nonparametric methods, including the feedforward neural networks with regularization to ensure Lipschitz continuity \citep*{gouk2021regularisation} and the kernel-based methods with nonexpansive kernels \citep*{van2022training}.

Given the above assumptions, the asymptotic normality of the proposed GPI estimator follows immediately from the DML theory.
\begin{theorem}[Asymptotic Normality of the GPI Estimator]\label{asymp_text} \spacingset{1}
Under Assumptions~\ref{consistency}--\ref{reg_text}, the estimator $\hat{\tau}$ obtained from the influence function $\psi$ satisfies asymptotic normality:
$$
\frac{\sqrt{n}(\hat{\tau} - \tau)}{\sigma} \xrightarrow[]{d} \mathcal{N}(0, 1)
$$
where $\sigma^2 = \E[\psi(\cD_i; \tau, \boldf, \mu_1, \mu_0, \pi)^2].$
\end{theorem}
The proof is given in Appendix~S4.3.  In addition, we can consistently estimate the asymptotic variance using the plugin estimator based on Equation~\eqref{eq:psi}. 

\subsection{Practical implementation details}
\label{subsec:implementation}

We discuss some important practical implementation details.  First, to satisfy Assumption~\ref{det_dec}, researchers must choose a deep generative model that has the option of deterministic decoding.  Aside from appropriately setting a hyper-parameter, the assumption also implies that we should not use batches with LLMs, which may induce unknown correlations across observations.  The use of LLMs that have memory should also be avoided.

In addition, the effective implementation of the proposed estimation method requires a careful choice of dimension reduction strategy for the internal representation, as well as hyperparameter tuning for TarNet.  First, the internal representation $\bR_i$, which typically corresponds to the last hidden states of a deep generative model, is of high dimension.  Specifically, it is a matrix of size equal to the length of texts $\times$ the size of representation for each token, which is equal to 768 for BERT-base, 1024 for BERT-large and T5-3B, and 4096 for Llama3-8B.  In theory, we can directly incorporate this matrix in TarNet.  In practice, however, given the limited computational resources available to researchers, it is advisable to apply a pooling operation to reduce the dimensionality.

The choice of pooling strategy depends on the architecture of a deep generative model.  For example, in BERT, the first special classification token [CLS] contains all semantic information \citep{devlin_bert_2019}.  Thus, we could use the hidden states that correspond to this [CLS] token alone. In BART (Bidirectional and Auto-Regressive Transformers), the special token is added at the end, so researchers can extract the hidden states of the last token \citep{lewis_bart_2020}. In contrast, encoder-decoder models like T5 (Text-to-Text Transfer Transformer, \citealt{raffel_exploring_2023}) do not have such special tokens, and mean pooling is often applied (e.g., \citealt{ni_sentence-t5_2021}).

For decoder-only models, such as Llama, the autoregressive structure forces all tokens to pay attention only to the past tokens. Hence, we can use the hidden states of the last token. This pooling strategy is frequently used as an approximation of the input text representation in the literature (e.g.,\citealt{neelakantan2022text, ma2024fine, jiang2023scaling}). We show the validity of this approximation in a simulation study (Section~\ref{sec:simulation}).  Our experience shows that this approximation generally works well.  If the confounding features are deemed too complex to be adequately captured by the last token alone, researchers may consider a sensitivity analysis \citep{lin2024isolatedcausaleffectsnatural}.

Second, for TarNet, we must carefully choose hyperparameters such as the size and depth of layers, learning rate, and maximum epoch size.  Together, they determine the success of optimization and the quality of the deconfounder estimate. The dimension of the deconfounder should be sufficiently large to capture the confounding information, but not too large to violate the overlap assumption. The learning rate and the epoch size are also crucial, as the performance is highly dependent on the success of the optimization process.

A practical strategy is to try different hyperparameter values and select the one that minimizes the loss.  If the loss does not decrease within the first few epochs, the optimization has likely failed, and researchers should try different  hyperparameter values. The process can be automated with advanced hyperparameter optimization methods, such as Optuna \citep{akiba_optuna_2019}, that search the optimal hyperparameters efficiently by dynamically constructing the search space.

\subsection{Diagnostic tools}\label{subsec:diagnostic_tools}

The key assumption of the GPI methodology is the separability between the treatment and confounding features (Assumption~\ref{separability}). This assumption concerns the functional relationship between the treatment and confounding features: (i) the treatment feature $T_i$ is not a deterministic function of $\bU_i$, and (ii) the confounding feature $\bU_i$ is not a function of $T_i$. As we show in Lemma~\ref{overlap}, condition (i) implies the overlap assumption. Condition (ii) implies that the confounding feature $\bU_i$ is disentangled from the treatment feature $T_i$, so that $\P\bigl(\bU_i \mid \mathrm{do}(T_i = t)\bigr) = \P\bigl(\bU_i \bigr)$ for all $t \in \cT$. \citet{wang_desiderata_2022} show that disentanglement implies the independence of support, i.e., $\mathrm{supp}(T_i) \times \mathrm{supp}(\bU_i) = \mathrm{supp}(T_i, \bU_i)$, which is a necessary condition for positivity. Therefore, checking positivity is crucial for diagnosing the potential violation of separability assumption.

We propose two ways to diagnose the positivity assumption. The first way is to plot the distribution of the estimated propensity scores $\hat{\pi}(\hat{\boldf}(\bR_i)) = \widehat{\P(T_i = 1 \mid \hat{\boldf}(\bR_i))}$ and assess whether they are bounded away from 0 and 1. If some estimated scores are too close to 0 or 1, one may trim them \citep{crump2009dealing}, clip them \citep{dorn2025much}, or use overlap weights \citep{li_addressing_2019}.

Another way to diagnose the potential violation of positivity is to compute the independence-of-support score (IOSS) proposed by \cite{wang_desiderata_2022}.  We use IOSS to measure the dependence of support between the deconfounder and the treatment variable.
After standardization, IOSS lies in $[0,1]$ and can be interpreted as the fraction of the standardized range by which the support of the two variables would need to be shifted in order to achieve perfect overlap.

The separability assumption also requires that the treatment feature can be manipulated independently of the confounding features. This implication is challenging to verify statistically. One practical approach is to instruct either an LLM or human to alter only the treatment feature while keeping the remaining aspects of the original text unchanged. The GPI methodology can then be applied to estimate treatment effects. If the manipulation is successful, the resulting treatment effect estimates should closely match those based on the original data. A limitation of this approach is that one needs to collect outcome data for the altered texts. Nevertheless, this provides a direct test of a core component of the separability assumption.

\section{Simulation Studies}
\label{sec:simulation}

We conduct simulation studies to evaluate the empirical performance of the GPI estimator and compare it to existing estimators.  

\subsection{Simulation setup}

We use the candidate profile experiment introduced in Section~\ref{sec:example} to make the simulation setup realistic.  We first generate candidate biographies using an open source LLM, Llama3--8B, that is fine-tuned for instruction-based prompt generations with 8 billion parameters.  We ask the model to create a biography of politicians under a hypothetical name using the system and user level prompts shown in Table~S2 of Appendix~S1.  We create hypothetical names by randomly drawing a surname and a first/middle name with replacement from the original corpus of \cite{fong_discovery_2016}.  The total number of biographies in our sample is 4,000.

We also examine the performance of the GPI methodology with the text reuse approach.  To do this, we instruct the same Llama3 model to exactly repeat each generated biography.  This allows us to compare our two approaches using the same set of texts.

After generating candidate biographies, we label them based on treatment and confounding features of interest. We consider two scenarios: the first is designed to adhere to the separability assumption (Assumption~\ref{separability}), which is our key assumption, while the second is likely to violate it.  
For the first scenario, we use the candidate's military background as the treatment variable. A biography is assigned to the treatment group if it contains at least one of the following keywords: ``military,'' ``veteran,'' or ``army.''

For confounding features, we first consider a combined topic of politics and education, denoted as $h_1(\bX_i)$ where $h_1$ is a complex function of the treatment object.  We employ a widely-used embedding-based topic model, \texttt{BERTopic} \citep{grootendorst_bertopic_2022}, and then assign $h_1(\bX_i) = 1$ if a generated biography is classified to a topic whose representative words include ``politics,'' ``student,'' ``college,'' ``elected,'' ``university,'' ``political,'' ``advocate,'' and ``education''.  As the second confounding feature, denoted by $h_2(\bX_i)$, we use the sentiment analysis module available in the Python \texttt{TextBlob} package, which yields a continuous sentiment score ranging from $-1$ to $1$.  Since the treatment feature is based on a set of specific keywords that are quite different from the confounding features, the separability assumption is likely to be satisfied under this scenario.

For the second scenario, we use two overlapping topics to define the treatment and confounding features so that the separability assumption is likely to be violated.  We again use topics obtained from \texttt{BERTopic}, and assign $T_i = 1$ if a generated biography is classified to a topic whose  representative words include ``college,'' ``political,'' ``elected,'' ``politics,'' ``student,'' ``senator,'' ``education,'' and ``legislative''.  For the confounding concept, we set $h_1(\bX_i) = 1$ if a biography is assigned to a topic whose representative words include ``college,'' ``political,'' ``senator,'' ``politics,'' ``elected,'' ``student,'' and ``career''.  Thus, although the treatment and confounding concepts are coded based on two different topics, they share many representative words, making it likely for the separability assumption to be violated.

Finally, we use the following linear model to generate the outcome variable,
$$
Y_i \ = \ \alpha_1 T_i + \alpha_2 T_i h_1(\bX_i) - \alpha_3 h_1(\bX_i) - \alpha_4 h_2(\bX_i) + \epsilon_i,
$$
where $\epsilon_i$ is the standard normal random variable.  Although the model may appear to be a relatively simple function of the treatment and confounding features, these variables themselves are complex functions of text.  Thus, in this simulation setting, inferring the ATE is not straightforward because it requires learning an accurate representation of confounding features.

To evaluate the performance of each estimator, we assume that researchers do not have access to the confounding features, $h_1(\bX_i)$ and $h_2(\bX_i)$.  We wish to infer the ATE, which is given by $\tau = \alpha_1 + \alpha_2 \E[h_1(\bX_i)]$ where we set $\alpha_1=\alpha_2=10$ throughout the simulations.  We consider the three scenarios: (1) weak confounding $\alpha_{3} = \alpha_{4} = 50$, (2) moderate confounding $\alpha_3 = \alpha_4 = 100$, (3) strong confounding $\alpha_3 = \alpha_4 = 1000$. Given the computational cost, our evaluation is conditional on a single set of generated biographies and its associated treatment and confounding variables.  Therefore, $\E[h_1(\bX_i)]$ is set to its sample mean, and the randomness comes only from the error term of the outcome model $\epsilon_i$.

\subsection{Estimators to be compared}

Once Llama3 generates all biographies, we extract an internal representation of each biography from the last hidden layer whose dimension is 4096 $\times$ the number of tokens contained in the biography.  For text reuse, we ask the LLM to repeat each generated text.  To compute the GPI estimator, we follow the discussion presented in Section~\ref{subsec:implementation} and use the representation of the last token, yielding a 4096 dimensional vector of internal representation for each candidate biography.

Our neural network architecture uses one linear layer for the deconfounder $\boldf(\bR_i)$ whose output dimension is 2048. Similarly, we utilize two consecutive linear layers for the potential outcome models whose output dimensions are 500 and 1. We apply ReLU as an activation function between each layer, and also use a dropout rate of 0.15 to prevent overfitting.  We select the neural network architecture, including dropout rates, layer depth, and dimension of each layer, based on additional hyperparameter tuning under the weak confounding setting. While we do not conduct hyperparameter tuning for each setting separately, we find that minor changes in these hyperparameters do not significantly affect the value of the loss function. 

We use 40\% of our data as a training set, 10\% as a validation set, and the remaining 50\% is used to estimate downstream causal effects. For optimization, we set the batch size to 32 and train the model for 500 epochs, and use early stopping to prevent overfitting.  Specifically, we stop training if the estimated loss does not improve for more than 15 epochs. Since the performance of the methodology can depend on the choice of hyperparameters, especially learning rates, we select learning rates using an automated hyperparameter tuning package \texttt{Optuna}. We do not include the size of the deconfounder as a hyperparameter to be tuned because tuning it for the BERT-based methods is computationally too expensive. We also do not observe any substantial difference in performance when varying the dimensions of the deconfounder for the GPI methodology.  Once the outcome model is fitted, we estimate the propensity score using a random forest classifier with the estimated deconfounder, using the \texttt{scikit-learn} library.  

We evaluate the performance of the GPI estimator in comparison to the following three estimators. First, as a baseline, we use the difference-in-means estimator, which makes no adjustment for confounding. Second, we implement the two existing approaches that estimate the vectorized representation of texts using the BERT embedding $\bm{b}(\cdot)$; \cite{pryzant_causal_2021} estimates nuisance functions with TarNet and calculates the causal effect using outcome models, while \cite{gui_causal_2023} uses the same outcome models as \cite{pryzant_causal_2021} but applies DML with the estimated propensity score. Importantly, both of these approaches are based on the following loss function,
\begin{equation}
    \frac{\lambda}{n}\sum_{i = 1}^n \left\{Y_i - Q_{T_i}(\bm{b}(\bX_i))\right\}^2 -\frac{\alpha}{n}\sum_{i = 1}^n \biggl\{  T_i\log g(\bm{b}(\bX_i)) + (1 - T_i) \log [ 1 - g(\bm{b}(\bX_i))] \biggr\}
    + \frac{1}{n}\sum_{i = 1}^n B(\bm{b}_{\mathrm{full}}(\bX_i)) \label{loss_pryzant}
\end{equation}
where $Q_{t}(\cdot)$ is the outcome model under $T_i = t$, $g(\cdot)$ is the treatment prediction, $B(\cdot)$ is the original BERT masked language loss for estimating vector representations of texts from the embedding $\bm{b}(\cdot)$, and $\alpha, \lambda \in \mathbb{R}$ are the hyperparameters \citep{veitch_adapting_2020}.  Only the vector representation of the first token $\bm{b}(\cdot)$ is used for prediction of outcome and treatment because it is the special token called the [CLS] token that contains the semantic information. On the other hand, for the masked language loss, the entire embedding $\bm{b}_{\mathrm{full}}(\cdot)$ is used.  The loss function given in Equation~\eqref{loss_pryzant} differs from our loss function (Equation~\eqref{loss_text_original}) in that in addition to the outcome model, it optimizes the representation learning from texts and the prediction of treatment.

After estimating nuisance functions, \cite{gui_causal_2023} proposes to estimate the propensity score model by treating $Q_1(\bm{b}(\bX_i))$ and $Q_0(\bm{b}(\bX_i))$ as covariates, i.e., $\Pr(T_i = 1 \mid \bX_i)=g(Q_1(\bm{b}(\bX_i)), Q_0(\bm{b}(\bX_i)))$. For propensity score estimation, we use a Gaussian Process, as recommended by \cite{gui_causal_2023}. We truncate the extreme values of the estimated propensity scores at 0.01 and 0.99 for DML with BERT, although such a truncation is not necessary for the GPI methodology.  For DML with BERT, this truncation occurs even when the separability assumption is satisfied (5\% of time for weak and moderate confounding, and 45\% for strong confounding).  Without truncation, the bias and RMSE are not computable for these methods. Finally, we follow the default hyperparameter used in the authors' original code and set $\alpha =\lambda = 1$ while tuning the other hyperparameters regarding the learning rate in the same way as done for the GPI estimator.

\subsection{Simulation results}
\label{subsec:sim-results}

\begin{figure}[t]
  \centering   \spacingset{1}
  \includegraphics[width=1.0\linewidth]{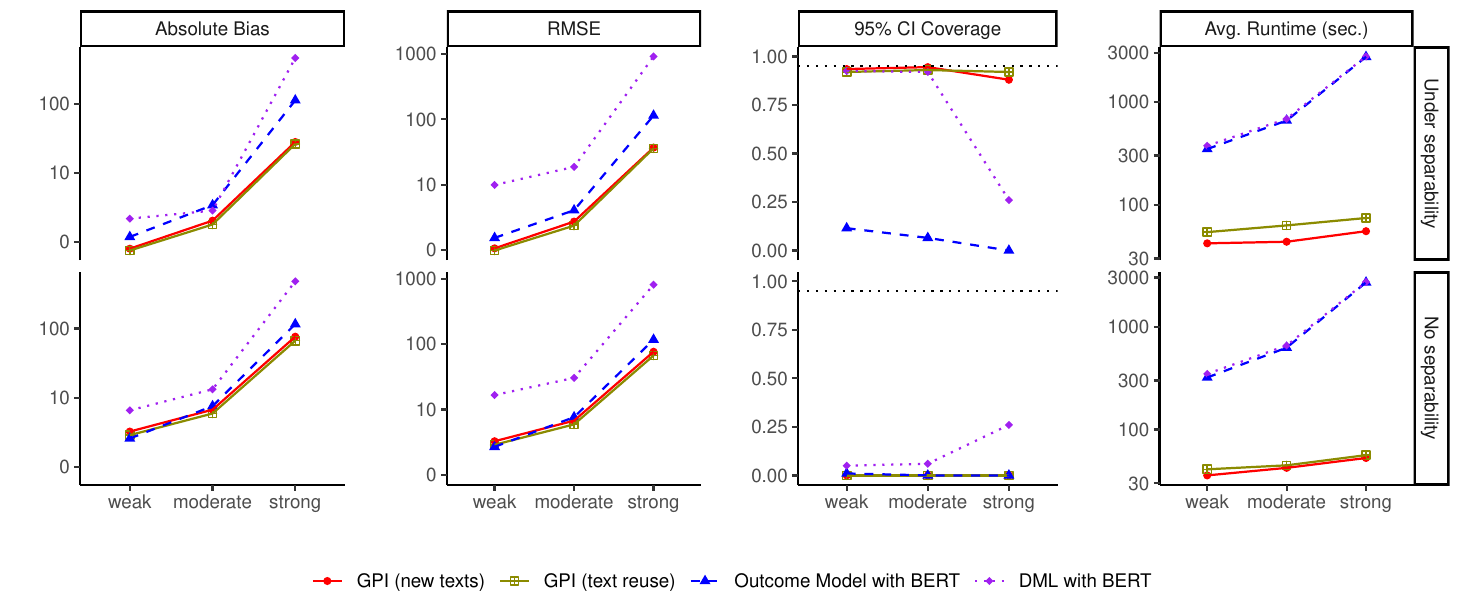}
    \caption{Performance of Five Estimators under Different Confounding Scenarios (Weak, Moderate, and Strong) under Separability (top row) and No Separability (bottom row). A red solid line represents the proposed GPI methodology for the new texts, while a dark yellow solid line represents that for the regenerated texts (text reuse).  For comparison, we also include outcome model with BERT (blue) and DML with BERT (purple).  The black horizontal dotted line in the 95\% Confidence Interval (CI) Coverage panel represents the nominal coverage of 95\%. }
    \label{all_results}
\end{figure}

Figure~\ref{all_results} graphically displays the results of our simulation studies while Appendix~S5 reports the corresponding numerical results.  The results are based on 200 Monte Carlo trials. We choose this relatively small number of trials because, unlike the GPI estimator, the BERT-based estimators are computationally intensive.  
 Appendix~S5 also presents the simulation results based on 1,000 Monte Carlo trials for the proposed estimator alone.  As expected, the results are qualitatively similar to those presented in this section.

We find that when the separability assumption holds (top row), the GPI estimators (red line for new texts and dark yellow for text reuse) exhibit a smaller bias and RMSE compared to all other estimators, with the 95\% confidence interval coverage closely matching the nominal rate. The performance differences are particularly striking in the strong confounding setting, where DML with BERT (purple) performs poorly, unlike the weak and moderate confounding scenarios. Additionally, outcome model with BERT (blue) has severe undercoverage across all simulation scenarios, whereas the confidence interval of DML with BERT breaks down only under the strong confounding scenario.  Lastly, the GPI estimators are more than ten times as computationally efficient as BERT-based estimators, when measured in terms of the average runtime.

In contrast, when the separability assumption is violated (bottom row), all estimators, including ours, perform poorly.  In this setting, both bias and RMSE grow as the strength of confounding increases.  As a result, the coverage of confidence intervals no longer approximates the nominal coverage rate even for the GPI estimator.

\begin{figure}[t]
  \centering  \spacingset{1}
  \includegraphics[width=1.0\linewidth]{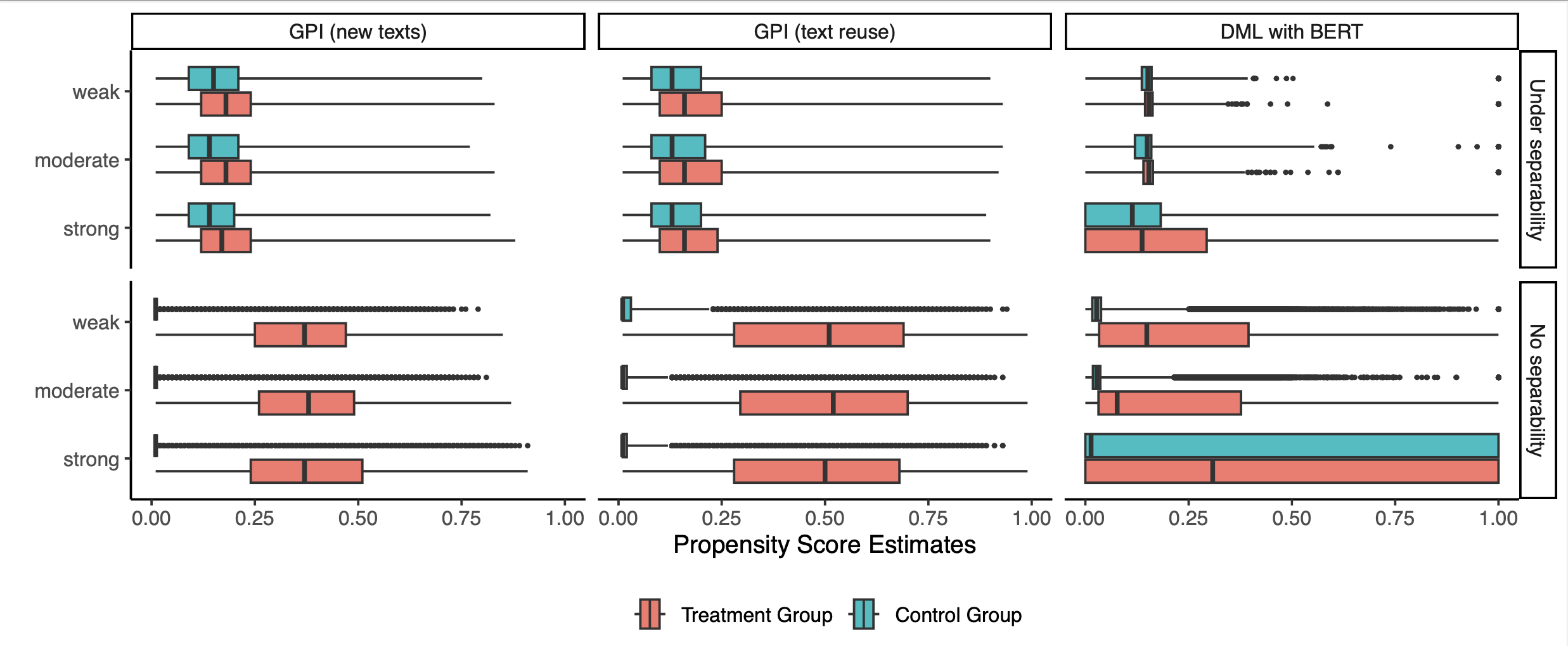}
    \caption{Distribution of the Estimated Propensity Scores for the Treatment (red) and Control (blue) Groups.  We present the results for the proposed GPI estimator (new texts in the left panels, and text reuse in the middle panels),  and DML with BERT (right) under the separability assumption (top) and no separability (bottom).  Under each scenario, we present the results based on three different strengths of confounding; weak, moderate, and strong.  For the proposed GPI methodology, the estimated propensity scores are distributed similarly across different confounding scenarios under the separability assumption.  In contrast, for DML with BERT, the distribution is heavily skewed right under the strong confounding scenario.  When the separability assumption is violated, both methods have extremely small estimated propensity scores for the control group.}
    \label{ps_density}
\end{figure}

We further examine the difference in performance between the GPI estimator and the DML with BERT. Figure~\ref{ps_density} shows the distribution of the estimated propensity scores without truncation (recall that truncation is not needed for the proposed methodology). For the GPI methodology, when the separability assumption is met (top panel), the distribution of estimated propensity scores is relatively symmetric regardless of the confounding strength and is similar between the treatment and control groups. Indeed, most observations have estimated propensity scores far from zero. This explains why the GPI estimator performs well in this scenario.

On the other hand, when the separability assumption is violated (bottom left and middle panels), the estimated propensity scores for the control group are heavily skewed to the right and close to zero, indicating that the overlap assumption is violated. This implies that the estimated propensity scores can help diagnose the potential violation of the separability assumption for our proposed method, which leads to complete entanglement between treatment and confounding features and causes the deconfounder to perfectly predict the treatment assignment. In contrast, DML with BERT yields a wide range of estimated propensity scores under the strong confounding settings (top right panel), in which the estimator performs poorly.  This pattern is observed even when the separability assumption is violated (bottom right panel).  The finding implies that, unlike the proposed estimator, extreme values of estimated propensity scores may not serve as a reliable diagnostic for DML with BERT.

\begin{figure}[t]
    \centering \spacingset{1}
    \includegraphics[width=0.9\linewidth]{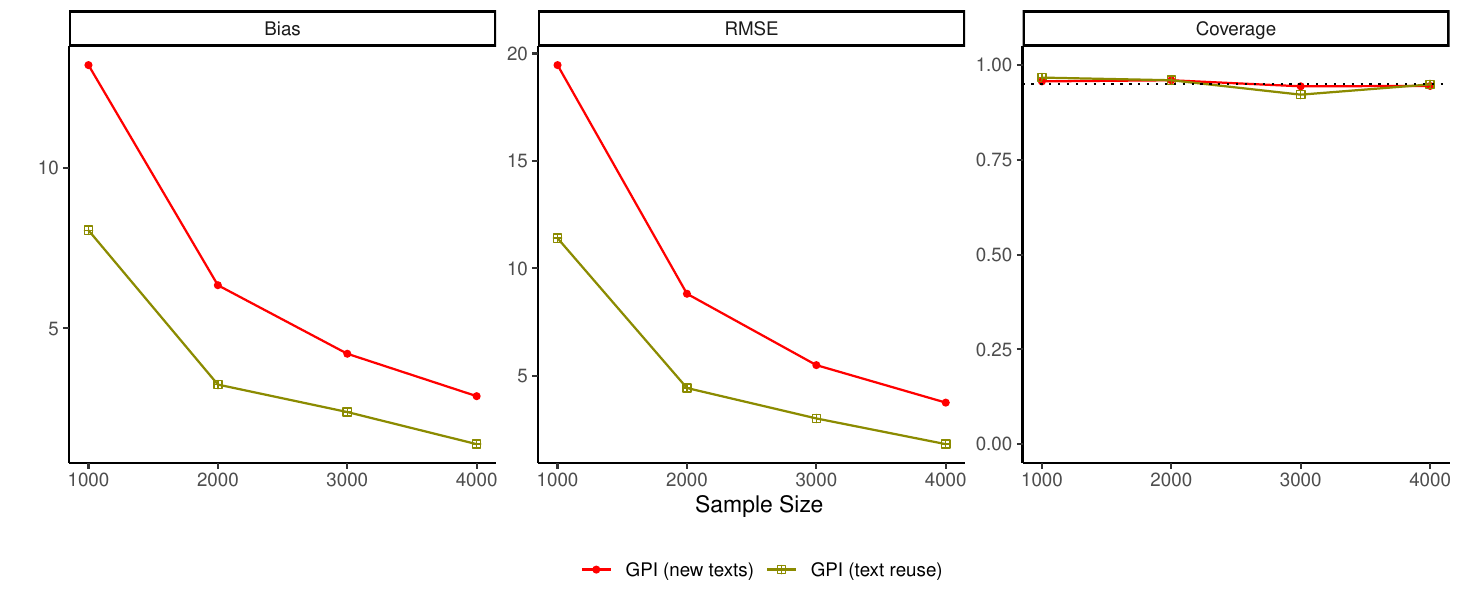}
    \caption{Performance of the GPI Estimator on the created texts (red) and the reused texts (blue) for Different Sample Size based on 1000 Monte Carlo trials. The data generating process is no interaction setting ($\alpha_1 = 10, \alpha_2 = 0, \alpha_3 = \alpha_4 = 100$). The black dotted line in the coverage panel represents the nominal coverage of 95\%. 
    }
    \label{sample_performance}
\end{figure}

Finally, we examine the performance of the GPI estimator for new texts and text reuse as we vary the sample size from 1,000 to 4,000 under the assumption of separability.  For this simulation, we use the moderate confounding setting without the interaction term ($\alpha_1 = 10, \alpha_2 = 0, \alpha_3 = \alpha_4 = 100$) so that the true ATE stays identical across sample sizes.  We conduct 1,000 Monte Carlo trials as we focus only on the GPI estimator, which is computationally efficient.  Figure~\ref{sample_performance} presents bias, RMSE, and coverage for each sample size. As expected, both bias and RMSE become smaller as the sample size increases.  We also find that the empirical coverage of the 95\% confidence intervals remains close to the nominal coverage even when the sample size is as small as 1000.  Interestingly, while bias is similar between new texts and text reuse, RMSE (hence variance) is substantially smaller for text reuse.  In sum, the proposed GPI methodology performs well in different sample sizes, provided that the separability assumption is satisfied.

\section{Empirical Analysis}
\label{sec:empirical}

Finally, we apply the proposed GPI methodology to human responses (rather than simulated data) from the candidate profile experiment of \cite{fong_discovery_2016} using pre-defined treatment coding. In the original experiment, the authors scraped 1,246 biographies of congressional candidates from Wikipedia, randomly assigned up to four biographies to 1,886 survey participants, and asked respondents how they felt about each candidate using the standard feeling thermometer ranging from 0 to 100, where a higher value indicates a more favorable evaluation.  The total number of observations we analyze is 5,291 after dropping some observations with empty texts.

While \cite{fong_discovery_2016} use a topic model to discover treatments from the data, we use the pre-defined treatment coding. As in simulation studies, we use a candidate's military background as the treatment variable by assigning a biography to the treatment group if it contains at least one of the following keywords: ``military,'' ``veteran,'' and ``army.''  According to this coding rule, only seven percent of the biographies (362 out of 5,291) are in the treatment group.

For causal effect estimation, we take the text reuse approach by regenerating each biography using Llama3-8B.  We then apply the proposed GPI methodology following the procedure described in Section~\ref{sec:methodology} with two-fold cross fitting.  For comparison, as in our simulation studies, we also apply the two existing BERT-based methods --- \cite{pryzant_causal_2021} (Outcome model with BERT) and \cite{gui_causal_2023} (DML with BERT).

\begin{table}[t]
    \centering \spacingset{1}
    \begin{tabular}{l.c..}
    \toprule
    Methods & \multicolumn{1}{c}{ATE Estimates} & \multicolumn{1}{c}{95\% Confidence Interval} & \multicolumn{1}{c}{IOSS} & \multicolumn{1}{c}{Runtime  (sec.)} \\
    \midrule
    GPI  (reuse) & 4.852 & [1.902, \hspace{.04in} 8.580] & 0.10 & 62.3 \\
    Outcome model with BERT & -4.277 & [$-$4.312, $-$4.241] & \multirow{2}{*}{\makebox[0pt][l]{\hspace*{-.08in}0.41}} & 5914.0  \\
    DML with BERT &  45.708 & [33.730,  57.686] &  & 5986.2  \\
    \bottomrule
    \end{tabular}
    \caption{The Estimated Average Treatment Effect (ATE) for the Candidate Profile Experiment.}
    \label{experiment}
\end{table}

Table~\ref{experiment} presents the results. The analysis based on the proposed GPI methodology implies that military experience has a positive effect and is statistically significant.  This echoes with the finding of \cite{fong_discovery_2016} that the topic corresponding to military experience has a statistically significant positive association with a higher feeling thermometer score. In contrast, outcome model with BERT yields a negative and statistically significant estimate, while DML with BERT produces a positive estimate that is unreasonably large given the scale is between 0 and 100.

To diagnose the potential violation of positivity, we also compute IOSS for both the GPI and BERT-based methods. While GPI yields IOSS of 0.10, the BERT-based methods yield a much larger IOSS of 0.41. This suggests that the separability assumption is much more likely to be voilated for the BERT-based methods than GPI. Indeed, for the BERT-based methods, all 5291 observations have estimated propensity scores outside of the range of $[0.01, 0.99]$.

Lastly, as observed in our simulation studies, the runtime for GPI is around 100 times shorter than that of the two BERT-based estimators.

\section{Concluding Remarks}
\label{sec:conclusion}

In this paper, we demonstrate that the use of GenAI can significantly enhance the validity of causal inference with unstructured treatments, such as texts and images. We leverage GenAI to both efficiently produce a variety of treatments and precisely control confounding bias. By utilizing the true vector representation of generated texts, we avoid estimating such representation as done in the previous methods, leading to more efficient and robust causal effect estimation.

We formalize the conditions required for nonparametric identification, showing that the separability of treatment and confounding features plays an essential role. We also develop an estimation method based on a neural network architecture that mitigates the risk of positivity violation, a common problem of existing methods. Lastly, we extend the proposed GPI methods to the settings of perceived treatments, using an instrumental variables approach. Our simulation study shows that the GPI estimator outperforms existing methods.

Although we have focused on texts as treatments, our GPI approach can be extended to other types of unstructured data, such as images and videos.  For images, we expect the proposed GPI methodology to be directly applicable so long as the dimensionality of internal representation is relatively low.  For videos, both treatment and confounding features are likely to exhibit complex relationships due to the combination of audio and images with the additional temporal dimension.  Thus, the GPI methodology described here is not readily applicable.  Future work should consider how to leverage the internal representation of videos obtained from GenAI.

\spacingset{1.4}
\addcontentsline{toc}{section}{\refname}
\bibliography{references,my}

\newpage
\appendix

\setcounter{equation}{0}
\setcounter{figure}{0}
\setcounter{table}{0}
\setcounter{section}{0}
\renewcommand {\theequation} {S\arabic{equation}}
\renewcommand {\thefigure} {S\arabic{figure}}
\renewcommand {\thetable} {S\arabic{table}}
\renewcommand {\thesection} {S\arabic{section}}

\begin{center}
  {\bf \Huge Supplementary Appendix}
\end{center}

\section{Examples of Candidate Biography and Prompt}
\label{app:prompt}

\begin{table}[h] 
    \centering \spacingset{1} \small
    \begin{tabular}{|p{15cm}|} 
       \hline
       \textbf{Candidate biography with military background} \\
        Anthony Higgins was born in Red Lion Hundred in New Castle County, Delaware. He attended Newark Academy and Delaware College, and graduated from Yale College in 1861, where he was a member of Skull and Bones. After studying law at the Harvard Law School, he was admitted to the bar in 1864 and began practice in Wilmington, Delaware. He also served for a time in the United States Army in 1864.\\ \\
        \hline
      \textbf{Candidate biography without military background} \\
        Benjamin Tappan was born in Northampton, Massachusetts, the second child and oldest son of Benjamin Tappan and Sarah (Homes) Tappan, who was a grandniece of Benjamin Franklin. Two of his younger brothers were abolitionists Arthur Tappan and Lewis Tappan. He attended the public schools in Northampton and traveled to the West Indies in his youth. He apprenticed as a printer and engraver, also studying painting with Gilbert Stuart. He read law to be admitted to the bar in Hartford, Connecticut, in 1799. Later that year, he moved to the Connecticut Western Reserve and founded what is now Ravenna, Ohio, laying out the original village in 1808. He married, March 20, 1801, Nancy Wright, sister of John C. Wright (congressman), afterwards a United States House of Representatives from Ohio. They had one son, Benjamin, born in 1812.\\
        \hline
    \end{tabular}
    \caption{Two examples of candidate biographies with and without military background from the experiment conducted by \cite{fong_discovery_2016}}
    \label{actual_bio}
\end{table}

\begin{table}[p]
    \centering \spacingset{1}
    \begin{tabular}{|p{7.75cm}|p{7.75cm}|}
       \hline
        \textbf{Prompts} & \textbf{Generated candidate biography} \\
        \hline
        \textbf{(System)}: You are a text generator who always produces a biography of the instructed person. \newline \textbf{(User)}:
        Create a biography of an American politician named Nathaniel C. Gilchrist & Nathaniel C. Gilchrist was an American politician who served as a member of the United States House of Representatives from Maryland's 4th congressional district from 1895 to 1899. Born on November 24, 1847, in Baltimore, Maryland, Gilchrist was the son of a prominent family of merchants and shipbuilders. He was educated at the Baltimore City College and later attended the University of Maryland, where he studied law. After graduating, he was admitted to the bar and began practicing law in Baltimore. \\
        \\
        \hline
        \textbf{(System)}: You are a text generator who just repeats the input text.\newline \textbf{(User)}: Higgins was born in Red Lion Hundred in New Castle County, Delaware. He attended Newark Academy and Delaware College, and graduated from Yale College in 1861, where he was a member of Skull and Bones. After studying law at the Harvard Law School, he was admitted to the bar in 1864 and began practice in Wilmington, Delaware. He also served for a time in the United States Army in 1864. & Higgins was born in Red Lion Hundred in New Castle County, Delaware. He attended Newark Academy and Delaware College, and graduated from Yale College in 1861, where he was a member of Skull and Bones. After studying law at the Harvard Law School, he was admitted to the bar in 1864 and began practice in Wilmington, Delaware. He also served for a time in the United States Army in 1864.\\
        \hline
    \end{tabular}
    \caption{Two examples of generated candidate biographies with Llama~3. The system-level input ({\bf System}) defines the type of tasks to be performed, whereas the user-level input ({\bf User}) defines a specific task to be performed.}
    \label{actual_text}
\end{table}

\newpage
\section{Illustrative Examples of the Separability Assumption}\label{separability_examples} 

In this Appendix section, we present illustrative example corpora of the separability assumption (Assumption~\ref{separability}).  We present two examples, one in which the assumption is satisfied and the other where it is violated. 

\subsection{Example corpus where the separability assumption is satisfied}

Table~\ref{corpus_separability} presents an illustrative example corpus, in which the separability assumption is satisfied.  In this toy example, the treatment feature \(T_i\in\{0,1\}\) equals 1 if a text contains a male pronoun (i.e., ``he,'' ``him,'' ``his'') and 0 otherwise. The confounding feature \(U_i\in\{0,1\}\) equals 1 if the sentence contains a ``lawyer'' or ``doctor'' and 0 otherwise. In this small corpus, \(T_i\) and \(U_i\) are associated: \(\Pr(T_i=1\mid U_i=1)=\tfrac{2}{3}\) and \(\Pr(T_i=1\mid U_i=0)=\tfrac{1}{3}\). The separability assumption nevertheless holds: it is possible to change the value of \(T_i\) by swapping male and female pronouns while leaving \(U_i\) unchanged.

\begin{table}[h]
  \centering
  \begin{tabular}{|l|c|c|l|l| }
    \hline
    \textbf{Original Text} & $T_i$ & $U_i$ & \textbf{Text with} $T_i = 1$ & \textbf{Text with} $T_i = 0$ \\
    \hline
    He is a lawyer. & 1 & 1 & He is a lawyer. & \underline{She} is a lawyer. \\
    She is a nurse. & 0 & 0 & \underline{He} is a nurse. & She is a nurse. \\
    He writes a book. & 1 & 0 & He writes a book. & \underline{She} writes a book. \\
    He is a doctor. & 1 & 1 & He is a doctor. & \underline{She} is a doctor. \\
    She gets married. & 0 & 0 & \underline{He} gets married. & She gets married. \\
    She is a doctor. & 0 & 1 & \underline{He} is a doctor. & She is a doctor. \\
    \hline
  \end{tabular} \spacingset{1}
  \caption{An example corpus where the separability assumption is satisfied. An underline represents a change made to the text when the treatment feature is altered.}
  \label{corpus_separability}
\end{table}

\subsection{Example corpus where the separability assumption is violated}

Table~\ref{corpus_no_separability} presents an illustrative corpus, in which the separability assumption is violated. Here, \(T_i=1\) if the sentence contains the honorific ``Mr.'' and \(U_i=1\) if it contains a male pronoun (``he,'' ``him,'' ``his''). In this setting, the editing operation that changes \(T_i\) from 0 to 1 (or vice versa) requires changing pronouns to maintain grammatical coherence. Consequently, \(U_i\) can change when \(T_i\) is altered, violating separability. For instance, in the first row, switching \(T_i\) from 0 to 1 changes \(U_i\) from 0 to 1.

\begin{table}[h]
  \centering
  \begin{tabular}{|p{5cm}|c|c|p{5cm}|p{5.2cm}| }
    \hline
    \textbf{Original Text} & $T_i$ & $U_i$ & \textbf{Text with} $T_i = 1$ & \textbf{Text with} $T_i = 0$ \\
    \hline
    Mrs. Park loves her children. & 0 & 0 & \underline{Mr.} Park loves \underline{his} children. & Mrs. Park loves her children. \\
    Mr. Lee met with the team. & 1 & 0 & Mr. Lee met with the team. & \underline{Mrs.} Lee met with the team. \\
    Mr. Zhou said he would pay. & 1 & 1 & Mr. Zhou said he would pay. & \underline{Mrs.} Zhou said \underline{she} would pay. \\
    Mrs. Li met him. & 0 & 1 & \underline{Mr.} Li met him. & Mrs. Li met him. \\
    \hline
  \end{tabular} \spacingset{1}
  \caption{A example corpus where the separability assumption is violated. An underline represents a change made to the text when the treatment feature is altered.}
  \label{corpus_no_separability}
\end{table}

\section{Instrumental Variable Approach to the Perceived Treatment Feature}
\label{sec:instrument}

The methodology described in the previous section enables the estimation of the average causal effect of the treatment feature, which is assumed to be a deterministic function of the treatment object.  In some cases, however, researchers may be interested in the causal effect of {\it perceived} treatment feature, which may not necessarily coincide with the treatment feature itself.  In addition, the perception of the same treatment feature may vary across respondents.  For example, in our application, respondents may disagree as to what constitutes a military background.

In this section, we extend the proposed methodology to the setting, in which the treatment feature is used as an instrumental variable for the perceived treatment feature.  As before, we describe the required assumptions, establish nonparametric identification, and propose estimation and inference strategies.

\subsection{Assumptions and causal quantity of interest}

We consider the same setting as in Section~\ref{sec:methodology} except that we observe the perceived treatment feature $\wT_i \in \{0,1\}$, which may not equal the treatment feature itself, i.e., $\wT_i \ne T_i$ for some $i$.  We assume that a respondent's perceived treatment feature is a function of the treatment and confounding features of the assigned treatment object.  We assume that both perceived treatment $\wT_i$ and original treatment $T_i$ are observed. 
\begin{assumption}[Perceived Treatment Feature] \label{perceived} \spacingset{1} The perceived treatment feature $\wT_i \in \{0,1\}$ is a function of treatment and confounding features, i.e.,
  $$\wT_i \ = \ \wT_i(T_i, \bU_i),$$
  where $\wT_i(t, \bu)$ is the potential value of the perceived treatment feature when the treatment variable $T_i$ is equal to $t \in \{0,1\}$ and the confounding variables $\bU_i$ equal $\bu \in \cU$.  
\end{assumption}

Importantly, under Assumption~\ref{perceived}, different respondents may perceive the same treatment feature differently.  In addition, it is also possible for confounding features to affect the perceived treatment feature.  In practice, researchers may measure the perceived treatment feature by asking respondents directly.  However, doing so may lead to the so-called {\it priming bias}, in which the act of asking this question itself draws a respondent's attention to the treatment feature and confounds the causal effect of interest.  To avoid this bias, researchers may measure the perceived treatment feature after the outcome variable is realized.  Addressing this methodological issue is beyond the scope of this paper, but interested readers should consult a recent literature on the topic \citep[see e.g.,][]{montgomery_how_2018, aronow_note_2019, klar_studying_2020, blackwell_priming_2023}.

To identify the causal effect of the perceived treatment feature, we use the treatment feature as an instrumental variable.  To do this, we define the potential outcome as a function of the perceived treatment feature, and the treatment and confounding features.  Formally, we replace Assumption~\ref{separability} with the following assumption while maintaining the same separability between the treatment feature and the confounding features.
\begin{assumption}[Separability with the Perceived Treatment Feature] \label{separability_iv} \spacingset{1} The potential outcome is a function of the perceived treatment $\Tilde{T}_i$, the treatment features of interest $T_i$, and the confounding features $\bU_i$. That is, for any given $\bx \in \cX$ and all $i$, we have:
$$
Y_i(\bx) = Y_i(\wT_i(g_{T}(\bx), \bg_{\bU}(\bx)), g_{T}(\bx), \bg_{\bU}(\bx))
$$
where $\wT_i(g_{T}(\bx), \bg_{\bU}(\bx))\in \{0,1\}$ is the perceived treatment feature, $g_{T}(\bx) \in \{0,1\}$, and $\bg_{\bU}(\bx) \in \cU$.  In addition, $g_T$ and $\bg_{\bU}$ are separable in the same sense as Assumption~\ref{separability}.
\end{assumption}

Lastly, we adopt the standard instrumental variable assumptions in current settings \citep{imbens_identification_1994}.  First, we assume monotonicity; the existence of treatment feature makes it no less likely for a respondent to perceive it as such. Second, we assume an exclusion restriction; the treatment feature only affects the outcome through the perceived treatment feature. Both assumptions are made while keeping the confounding features constant.  We formally state these assumptions here.  
\begin{assumption}[Validity of the Instrumental Variable] \label{iv} \spacingset{1}
  We make the following instrumental variable assumptions:
  \begin{enumerate}[label=(\alph*)]
\item (Monotonicity) \label{monotonicity} For any $\bu \in \cU$, we have: 
  $$\wT_i(1, \bu) \geq \wT_i(0, \bu) \quad \text{and} \quad \P(\wT_i(1, \bu) = 1) > \P(\wT_i(0, \bu) = 1).$$

\item (Exclusion Restriction) \label{exclusion} For any $\tilde{t} \in \{0,1\}$, $\bu \in \cU$, and $i=1,2,\ldots,n$, we have:
$$
Y_i(\tilde{t}, 1, \bu) \ = \ Y_i(\tilde{t}, 0, \bu) \ = \  Y_i(\tilde{t}, \bu) .
$$
  \end{enumerate}
\end{assumption}
In many practical settings, the monotonicity assumption is reasonable.  In our application, for example, if there is no military background in a candidate biography, a respondent should not notice the presence of this treatment feature.  Exclusion restriction, however, may not be credible in some cases because it is possible for a respondent to be influenced by the treatment feature without noticing it.

Under this setup, we are interested in estimating the local average treatment effect (LATE) of the perceived treatment feature among the respondents who notice the presence of the treatment feature only when the treatment object actually contains such a feature. We define this LATE as follows:
\begin{align}
    \beta \ := \ \mathbb{E}[Y_i(1,\bU_i) - Y_i(0,\bU_i)\mid \wT_i(1,\bU_i) = 1, \wT_i(0,\bU_i)=0], \label{late}
\end{align}
where the first input of the potential outcome is the perceived treatment feature instead of the treatment feature, i.e., $Y_i(\wT_i = \tilde{t}, \bU_i = \bu)$. 

\subsection{Nonparametric identification}
\label{subsec:identification_iv}

\begin{figure}[t]
\centering \spacingset{1}
\begin{tikzpicture}
\node[text centered] (l) {$\bP$};
\node[right = 1.5 of l, text centered] (l2) {$\bR$};
\node[right = 1 of l2, text centered] (l3) {$\bh_{\bgamma}(\bR)$}; 
\node[right = 1.5 of l3, text centered] (r) {$\bX$};
\node[above right = 1 of r, text centered] (c) {$T = g_T(\bX)$};
\node[right = 1 of c, text centered] (t) {$\wT$};
\node[below right= 1.0 of r, text centered] (p) {$\bU = \bg_{\bU}(\bX)$};
\node[right= 6 of r, text centered] (y) {$Y$};
\node[rectangle, draw, minimum width = 4.25cm, minimum height = 1.5cm] (z) at (3.25,0) {}; 
\node[rectangle, minimum width = 4.25cm, minimum height = 1.5cm] (z) at (3.25,1) {Deep generative model}; 

\draw[-Stealth] (l) --  (l2);
\draw[-Stealth, double, color = red] (l2) --  (l3);
\draw[-Stealth, double, color = red] (l3) --  (r);
\draw[-Stealth, double, color = red] (r) --  (c);
\draw[-Stealth, double, color = red] (r) --  (p);
\draw[-Stealth] (p) --  (t);
\draw[-Stealth] (c) --  (t);
\draw[-Stealth] (p) --  (y);
\draw[dashed, <->] (t.east) to [out=60,in=40] (y.north);
\draw[-Stealth] (t) --  (y);
\end{tikzpicture}
\caption{Directed Acyclic Graph (DAG) of the Assumed Data Generating Process with the Perceived Treatment Feature.  This DAG is identical to that of Figure~\ref{DAG} except that the perceived treatment feature $\wT$ is added.  The perceived treatment feature may be affected by the treatment feature $T$ and/or the confounding features $\bU$.  There may also be unobserved confounding variables that affect both the perceived treatment feature and the outcome $Y$.  An arrow with red double lines represents a deterministic causal relation while an arrow with a single line indicates a possibly stochastic relationship.}
\label{dag_percep}
\end{figure}

We extend our nonparametric identification result obtained in Section~\ref{subsec:identification} to the instrumental variable setting.  Figure~\ref{dag_percep} summarizes the assumed data generation process with the perceived treatment feature.  The absence of direct arrow from the treatment feature $T$ into $Y$ encodes exclusion restriction (Assumption~\ref{iv}\ref{exclusion}).  In addition, we allow for the possible existence of unobserved confounders between the perceived treatment feature and the outcome, indicated by the dotted line in the DAG.

The following theorem establishes the nonparametric identification of the LATE defined in Equation~\eqref{late}.  As in the case of ATE (see Theorem~\ref{iden_det}), identification is achieved by adjusting for the deconfounder $\boldf(\bR_i)$ and using the treatment feature $T_i$ as an instrument for the perceived treatment feature.  Similarly to the ATE case, the deconfounder satisfies the conditional independence relation $\{Y_i, \wT_i\} \indep \bR_i \mid T_i = t, \boldf(\bR_i)$.  The difference is that the inner representation $\bR_i$ is now independent of the perceived treatment feature as well as the outcome after conditioning on the treatment feature and the deconfounder.  Finally, we emphasize that like Theorem~\ref{iden_det}, this result does not require the deconfounder to be unique. 
\begin{theorem}[Nonparametric Identification of the LATE]\label{iden_perc} \spacingset{1} Under Assumptions~\ref{consistency}--\ref{confounder},~\ref{det_dec},~\ref{perceived}--\ref{iv}, there exists a deconfounder function $\boldf: \cR \to \cQ$ with $d_Q=\dim(\cQ) \le d_R = \dim(\cR)$ that satisfies the following conditional independence relation:
\begin{align*}
    \{Y_i, \wT_i\} \indep \bR_i \mid T_i = t, \boldf(\bR_i) = \bq,
\end{align*}
for all $\bq \in \cQ$ and $t=0,1$.  In addition, the treatment feature and the deconfounder are separable.   Then, by adjusting for such a deconfounder, we can uniquely and nonparametrically identify the local average treatment effect (LATE) defined in Equation~\eqref{late} as:
\begin{align*}
\beta 
& \ = \ 
\frac{\ \int_{\cR} \E[Y_i \mid T_i = 1, 
    \boldf(\bR_i)] - \E[Y_i \mid T_i = 0, 
    \boldf(\bR_i)] dF(\bR_i)}{\int_{\cR} \E[\wT_i  \mid T_i = 1, \boldf(\bR_i)] - \E[\wT_i \mid T_i = 0, \boldf(\bR_i)] dF(\bR_i)}.
\end{align*}
\end{theorem}
The proof is given in Appendix~\ref{proof_iden_perc}. 

\subsection{Estimation and inference}

\begin{figure}[t]
    \centering \spacingset{1}
\begin{tikzpicture}[
    node distance=0.5cm and 0.5cm,
    box/.style={draw, rectangle, minimum height=0.5cm, minimum width=0.5cm},
    func/.style={draw, rectangle, minimum size=0.5cm},
    ]
    
    \node[box] (z1) {$\bR$};
    \node[box, right = 1.5 of z1, fill = lightgray] (z3) {$\boldf(\bR; \blambda)$};
    \node[box, above right= 2.0 of z3, fill = lightgray] (t13) {$\mu_1(\boldf(\bR; \blambda); \btheta_1)$};
    \node[box, below right= 2.0 of z3, fill = lightgray] (t03) {$\mu_0(\boldf(\bR; \blambda); \btheta_0)$};
    \node[box, below = 0.4 of t13, fill = lightgray] (h11) {$m_1(\boldf(\bR; \blambda); \bzeta_1)$};
    \node[box, above = 0.4 of t03, fill = lightgray] (h01) {$m_0(\boldf(\bR; \blambda); \bzeta_0)$};

    \node[box, right=1.5 of t13] (t13label) {$Y_i\mid T_i = 1$}; 
    \node[box, right=1.5 of t03] (t03label) {$Y_i\mid T_i = 0$}; 
    \node[box, right=1.5 of h11] (h11label) {$\widetilde{T}_i \mid T_i=1$};  
    \node[box, right=1.5 of h01] (h01label) {$\widetilde{T}_i \mid T_i=0$};   
    
    \draw[->, line width= 1] (z1) --  (z3); 
    \draw[->, line width= 1] (z3) to  (t03.west);
    \draw[->, line width= 1] (z3) to  (t13.west);
    \draw[->, line width= 1] (z3) to  (h01.west);
    \draw[->, line width= 1] (z3) to  (h11.west);
    \draw[->, line width= 1] (t13) --  (t13label); 
    \draw[->, line width= 1] (t03) --  (t03label); 
    \draw[->, line width= 1] (h11) --  (h11label); 
    \draw[->, line width= 1] (h01) --  (h01label); 
\end{tikzpicture}
\caption{Diagram Illustrating the Proposed Model Architecture with the Instrumental Variable. The proposed model takes an internal representation of a treatment object $\bR_i$ as an input, and finds a deconfounder $\boldf(\bR_i)$, which is a lower-dimensional representation of $\bR_i$, and then use it to predict the conditional expectations of the outcome $\mu_{t}(\boldf(\bR_i)) := \E[Y_i \mid T_i = t,\boldf(\bR_i)]$ and the perceived treatment feature $m_{t}(\boldf(\bR_i)) := \E[\wT_i \mid T_i = t, \boldf(\bR_i)]$ under each treatment arm $t$.} \label{model_iv}
\end{figure}

Next, we extend the estimation and inference approaches developed in Section~\ref{subsec:estimation} to the instrumental variable setting.  The main difference is that we additionally model the conditional expectation of the perceived treatment feature given the treatment feature and deconfounder,
\begin{equation*}
  m_t(\boldf(\bR_i)) \ := \ \E[\wT_i \mid T_i = t, \boldf(\bR_i)],
\end{equation*}
for $t \in \{0,1\}$. Figure~\ref{model_iv} presents the proposed neural network architecture that extends the diagram shown in Figure~\ref{tarnet} to the instrumental variable setting.  The loss function is given by,
\begin{equation}
  \begin{aligned}
    \{\hat\blambda, \hat\btheta,\hat\bzeta\} \ = \ \argmin_{\blambda, \btheta,\bzeta} &  \frac{1}{n} \sum_{i=1}^n \left[\left(Y_i - \mu_{T_i}(\boldf(\bR_i; \blambda); \btheta_{T_i})\right)^2+ \left(\wT_i - m_{T_i}(\boldf(\bR_i; \blambda); \bzeta_{T_i})\right)^2 \right]
  \end{aligned}
  \label{loss_perc}
\end{equation}

Given this neural network architecture, we again use the DML framework for estimation and inference.  The exact estimation procedure is described here for completeness. 
\begin{enumerate} \spacingset{1}
    \item Randomly partition the data into $K$ folds of equal size where the size of each fold is $n = N/K$.  The observation index is denoted by $I(i)\in \{1,\dots,K\}$ where $I(i)=k$ implies that the $i$th observation belongs to the $k$th fold.
    \item For each fold $k \in \{1, \cdots, K\}$, use observations with $I(i)\ne k$ as training data:
    \begin{enumerate}
    \item split the training data into two folds, $I_1^{(-k)}$ and $I_2^{(-k)}$ 
    \item obtain estimates of deconfounder and the conditional outcome function on the first fold, denoted by $$
      \begin{aligned}
        \hat\boldf^{(-k)}(\{\bR_i\}_{i \in I_1^{(-k)}}) & \  := \ \boldf(\{\bR_i\}_{i \in I_1^{(-k)}}; \hat\blambda^{(-k)}), \\
        \mu_t^{(-k)} & \ := \ \mu_t(\hat\boldf(\{\bR_i\}_{i \in I_1^{(-k)}}; \blambda^{(-k)}); \hat\btheta^{(-k)}),
\end{aligned}
$$ and that of the perceived treatment feature, denoted by
$$
\hat{m}_t^{(-k)}(\{\bR_i\}_{i \in I_1^{(-k)}}):=m_t (\boldf(\{\bR_i\}_{i \in I_1^{(-k)}}; \hat\blambda^{(-k)}); \hat\bzeta^{(-k)})$$
by solving the optimization problem given in Equation~\eqref{loss_perc}, and
    \item obtain an estimate of the propensity score given the estimated deconfounder on the second fold, denoted by $\hat{\pi}^{(-k)}(\hat\boldf^{(-k)}(\{\bR_i\}_{i \in I_2^{(-k)}})):=\hat\pi^{(-k)} (\boldf(\{\bR_i\}_{i \in I_2^{(-k)}}; \hat\blambda^{(-k)}))$. 
    \end{enumerate}
    \item Compute an LATE estimate $\hat{\beta}$ as a solution to:
        \begin{align*}
            \frac{1}{nK}\sum_{k = 1}^K  \sum_{i: I(i)=k} \phi(\widetilde\cD_i; \hat{\beta}, \hat{\boldf}^{(-k)}, \hat\mu^{(-k)}_1, \hat\mu^{(-k)}_0, \hat{m}^{(-k)}_1, \hat{m}^{(-k)}_0, \hat\pi^{(-k)}) \ = \ 0,
        \end{align*}
        where $\widetilde{\cD}_i :=\{Y_i, \wT_i, T_i, \bR_i\}$ and
        \begin{equation*}
        \begin{aligned}
          & \phi(\widetilde{\cD}_i; \beta, \boldf, \mu_1, \mu_0, m_1, m_0, \pi) \\
          \ = \ & \frac{T_i \{Y_i - \mu_1(\boldf(\bR_i))\}}{\pi(\boldf(\bR_i)} - \frac{( 1- T_i) \{Y_i - \mu_0(\boldf(\bR_i))\}}{1 - \pi(\boldf(\bR_i))} + \mu_1(\boldf(\bR_i)) - \mu_0(\boldf(\bR_i)) \\
          & - \left[\frac{T_i \{\wT_i - m_1(\boldf(\bR_i))\}}{\pi(\boldf(\bR_i))} - \frac{( 1- T_i) \{\wT_i - m_0(\boldf(\bR_i))\}}{1 - \pi(\boldf(\bR_i))} + m_1( \boldf(\bR_i)) - m_0(\boldf(\bR_i)) \right] \cdot \beta.
        \end{aligned}
      \end{equation*}
      \end{enumerate}
      
Similar to the ATE case, we can establish the asymptotic property of this estimator. We first outline a set of additional regularity conditions required beyond Assumption~\ref{reg_text}. 
\begin{assumption}[Additional Regularity Conditions]\label{reg_perc} \spacingset{1}
Let $c_1$, $c_2$, and $q > 2$ be positive constants and $\delta_n$ be a sequence of positive constants approaching zero as the sample size $n$ increases.  Then, the following conditions hold:
  \begin{enumerate}[label=(\alph*)]
    \item (Primitive condition)
    \begin{align*}
       &  \E\left[\left\lvert \left(Y_i - \mu_{T_i}(\boldf(\bR_i))\right) - \beta \cdot \left( \widetilde T_i - m_{T_i}(\boldf(\bR_i))\right) \right \rvert^2\right]^{1/2} \geq c_2 
    \end{align*}
   \item (Perceived treatment model estimation)
    \begin{align*}
      & \E[m_1(\boldf(\bR_i)) - m_0(\boldf(\bR_i))] \geq c_2, \quad \E[|\hat{m}_{T_i}(\hat\boldf(\bR_i)) - m_{T_i}(\boldf(\bR_i))|^q]^{1/q} \le c_1, \\
      & \E[|\hat{m}_{T_i}(\hat\boldf(\bR_i)) - m_{T_i}(\boldf(\bR_i))|^2]^{1/2} \leq \delta_n n^{-1/4}.
     \end{align*}          
    \end{enumerate}
\end{assumption}
Together with Assumption~\ref{reg_text}, these regularity conditions are essentially equivalent to the assumptions required for DML inference on LATE \citep{chernozhukov_doubledebiased_2018}.

Under the above assumptions, the asymptotic normality of the proposed estimator can be established.
\begin{theorem}[Asymptotic Normality of Instrument Variable Estimator]\label{asymp_perc} \spacingset{1}
 Under Assumptions~\ref{consistency}--\ref{confounder},~\ref{det_dec}--\ref{reg_perc}, the estimator $\hat{\beta}$ obtained from the influence function $\phi$ satisfies asymptotic normality:
$$
\frac{\sqrt{n}(\hat{\beta} - \beta)}{\sigma} \xrightarrow[]{d} \mathcal{N}(0, 1)
$$
where
$
\sigma^2 = \E[\phi(\widetilde{\cD}_i; \beta, \boldf, \mu_1, \mu_0, m_1, m_0, \pi)^2] / \E[\gamma_1(\boldf(\bR_i)) - \gamma_0(\boldf(\bR_i))]^2. $
\end{theorem}
Proof is omitted given that, like Theorem~\ref{asymp_text}, the result follows immediately from the application of DML theory \citep{chernozhukov_doubledebiased_2018}.

\section{Proofs}

\subsection{Proof of Lemma \ref{overlap}} \label{proof_overlap}

We use proof by contradiction. Suppose that the overlap condition is not satisfied. That is, there exist $t \in \{0,1\}$ and $\bu \in \cU$ such that $\P(T_i = t\mid \bU_i = \bu)=0$.  This implies that under Assumptions~\ref{det_trt}~and~\ref{confounder}, there exist a deterministic function $\tilde{g}_T: \cU \rightarrow \{0,1\}$ and some $\bx \in \cX$ such that $t = \tilde{g}_T(\bu)=\tilde{g}_T(\bg_{\bU}(\bx))$.  This contradicts Assumption~\ref{separability}. \qed

\subsection{Proof of Theorem \ref{iden_det}}
\label{proof_iden_det}

Under a deep generative model of Definition~\ref{deep_use}, the distribution of $\bX_i$ only depends on $\bP_i$, and hence we have $Y_i(\bx) \indep \bX_i \mid \bP_i$. Together with Assumption~\ref{ignorability}, Lemma~4.3 of \cite{dawid_conditional_1979} implies $Y_i(\bx) \indep \bX_i$. Then, under Assumptions~\ref{det_trt}~and~\ref{separability}, we have:
\begin{equation}
  Y_i(t, \bU_i) \indep T_i \mid \bU_i. \label{eq:YtUindepTgivenU}
\end{equation}
Next, under
Assumptions~\ref{det_trt},~\ref{confounder},~and~\ref{det_dec},
$\bU_i$ is a deterministic function of $\bR_i$ such that we can
write $\bU_i = \boldf^\ast(\bR_i)$ for some function $\boldf^\ast: \cR \rightarrow \cU$.
Furthermore, Assumption~\ref{separability} implies
$Y_i \indep \bR_i \mid T_i = t, \boldf^\ast(\bR_i)=\bu$ and $0 < \Pr(T_i = t \mid \boldf^\ast(\bR_i) = \bu) < 1$ for all $t \in \{0, 1\}$ and $\bu \in \cU$.  Thus, we have:
\begin{align*}
    \P(Y_i(t, \bU_i) = y) & \ = \ \int_{\cU} \P(Y_i(t, \bU_i) = y \mid \bU_i)
                        dF(\bU_i) \\
                      & \ = \ \int_{\cU} \P(Y_i(t, \bU_i) = y \mid T_i
                        = t, \bU_i) dF(\bU_i) \\
                          & \ = \ \int_{\cU} \P(Y_i = y \mid T_i = t, 
                            \boldf^\ast(\bR_i)) dF(\boldf^\ast(\bR_i)), \\
                          & \ = \ \int_{\mathcal{R}} \P(Y_i = y \mid T_i = t, 
                            \boldf^\ast(\bR_i)) dF(\bR_i), 
\end{align*}
where the second equality follows from
Equation~\eqref{eq:YtUindepTgivenU} and Lemma~\ref{overlap},
and the third equality is due to Assumption~\ref{consistency}.
Finally, suppose there is another function $\boldf: \cR \rightarrow \cQ$, which
satisfies the conditional independence relation
$Y_i \indep \bR_i \mid T_i = t, \boldf(\bR_i)=q$ for all $q \in \cQ$ and is separable from the treatment feature.  Then,
\begin{align*}
  \int_{\mathcal{R}} \P(Y_i = y \mid T_i = t, 
  \boldf(\bR_i)) dF(\bR_i)
  \ = \ & \int_{\mathcal{R}} \P(Y_i = y \mid T_i = t, \boldf(\bR_i), \bR_i) dF(\bR_i) \\
  \ = \ & \int_{\mathcal{R}} \P(Y_i = y \mid T_i = t, \boldf^\ast(\bR_i), \bR_i) dF(\bR_i) \\
  \ = \ & \int_{\mathcal{R}} \P(Y_i = y \mid T_i = t, \boldf^\ast(\bR_i)) dF(\bR_i) 
\end{align*}
Thus, any function of $\bR_i$ that satisfies this conditional independence relation leads to the same identification formula for the marginal distribution of potential outcome. \qed

\subsection{Proof of Theorem~\ref{asymp_text}}\label{proof_asymp_text}

Assumptions~\ref{reg_text}\ref{reg_text_deconfounder}--\ref{reg_text_propensity} and the triangule inequality imply,
\begin{align}
  \E[|\hat{\pi}(\hat{\boldf}(\bR_i)) - \pi(\boldf(\bR_i))|^q]^{1/q}
  \ = \ &  \E[|\{\hat{\pi}(\hat{\boldf}(\bR_i)) - \hat{\pi}(\boldf(\bR_i))\} + \{\hat{\pi}(\boldf(\bR_i)) - \pi(\boldf(\bR_i))\}|^q]^{1/q} \nonumber \\
   \leq \ & \E[|\hat{\pi}(\hat{\boldf}(\bR_i)) - \hat{\pi}(\boldf(\bR_i))|^q]^{1/q} + c_1 \nonumber\\
   \leq \ &  L \cdot \E\left[\norm{\hat{\boldf}(\bR_i) - \boldf(\bR_i)}^q\right]^{1/q} + c_1 \nonumber \\
  = \ &  (L+1) c_1 \label{eq:bound_pi}
\end{align}
where $L$ is a Lipschitz constant.  Similarly, we can also show,
\begin{align}
  & \E\left[\left|\hat{\pi}(\hat{\boldf}(\bR_i)) - \pi(\boldf(\bR_i)) \right|^2 \right]^{1/2}\nonumber \\
  \leq \ &  L \cdot \E\left[ \norm{\hat{\boldf}(\bR_i) - \boldf(\bR_i)}^2 \right]^{1/2} 
    + \E\left[\left|\hat{\pi}(\boldf(\bR_i)) - \pi(\boldf(\bR_i))\right|^2\right]^{1/2}\nonumber \\
  \leq \ &  (L + 1) \delta_n n^{-1/4}. \label{eq:conv_pi}
\end{align}
Together with Assumption~\ref{reg_text}\ref{reg_text_outcome}, Equation~\eqref{eq:conv_pi} implies that there exists a sequence of positive constants $\delta_n'$ converging to zero as the sample size $n$ increases such that the following inequality holds,
\begin{align}
  \E[|\hat{\pi}(\hat{\boldf}(\bR_i)) - \pi(\boldf(\bR_i))|^2]^{1/2} \cdot \E[|\hat\mu_{T_i}(\hat\boldf(\bR_i)) - \mu_{T_i}(\boldf(\bR_i))|^2]^{1/2} \leq \delta_n' n^{-1/2}. \label{eq:conv_cross}
\end{align}
Thus, the standard regularity conditions of the DML theory \citep{chernozhukov_doubledebiased_2018} are satisfied for the estimated propensity score with the estimated deconfounder, i.e., $\hat{\pi}(\hat{\boldf}(\bR_i))$.  Finally, Assumptions~\ref{reg_text}\ref{reg_text_primitive}--\ref{reg_text_outcome} and Equations~\eqref{eq:bound_pi}~and~\eqref{eq:conv_cross} imply,
\begin{align*}
  \sqrt{n}\left(\hat{\tau} - \tau  \right) \xrightarrow[]{d} \mathcal{N}(0, \sigma^2)
\end{align*}
where $\sigma^2 = \E[\psi(\cD_i; \tau, \boldf, \eta_1, \eta_0, \pi)^2]$.

\qed

\subsection{Proof of Theorem \ref{iden_perc}}
\label{proof_iden_perc}

Under a deep generative model (Definition~\ref{deep_use}), the distribution of $\bX_i$ only depends on $\bP_i$, and hence we have $\{Y_i(\bx),\wT_i(\bx)\} \indep \bX_i \mid \bP_i$. Together with Assumption~\ref{ignorability}, Lemma~4.3 of \cite{dawid_conditional_1979} implies $\{Y_i(\bx),\wT_i(\bx)\} \indep \bX_i$. Under Assumptions~\ref{det_trt},~\ref{confounder},~\ref{perceived},~\ref{separability_iv},~and~\ref{iv}~\ref{exclusion}, we have, for any $t,\tilde{t} \in \{0, 1\}$ and $\bu \in \cU$:
\begin{equation}
    \{Y_i(\tilde{t}, \bu), \wT_i(t, \bu)\} \indep \{T_i, \bU_i\}. \label{eq:YwTindepTU}
\end{equation}
Then, for any $\bu \in \cU$,
\begin{align*}
    &\E[Y_i \mid T_i = 1, \bU_i = \bu] - \E[Y_i \mid T_i = 0, \bU_i = \bu]\\
   = \  & \E[Y_i(\wT_i(1, \bu), \bu) \mid T_i = 1, \bU_i = \bu] - \E[Y_i(\wT_i(0, \bu), \bu) \mid T_i = 0, \bU_i = \bu]\\
   = \  & \E[Y_i(\wT_i(1, \bu), \bu) - Y_i(\wT_i(0, \bu), \bu)]\\
   = \  & \E[Y_i(\wT_i(1, \bu), \bu) - Y_i(\wT_i(0, \bu),  \bu) \mid \wT_i(1, \bu) =1, \wT_i(0, \bu)= 0] \cdot \P(\wT_i(1, \bu) = 1, \wT_i(0, \bu) = 0 )\\
  =  \  & \E[Y_i(1, \bu) - Y_i(0, \bu) \mid \wT_i(1, \bu) =1, \wT_i(0, \bu)=0] \cdot \P(\wT_i(1, \bu) =1 , \wT_i(0, \bu) = 0),
\end{align*}
where the second equality follows from Equation~\eqref{eq:YwTindepTU}, the third equality is due to Assumption~\ref{iv}. 
We also have:
\begin{align*}
    \P(\wT_i(1, \bu) =1, \wT_i(0, \bu)=0) &= \E[\wT_i(1, \bu) - \wT_i(0, \bu)]\\
    &= \E[\wT_i(1, \bu)  \mid T_i = 1, \bU_i = \bu] - \E[\wT_i(0, \bu) \mid T_i = 0, \bU_i = \bu]\\
    &= \E[\wT_i  \mid T_i = 1, \bU_i = \bu] - \mathbb{E}[\wT_i \mid T_i = 1, \bU_i = \bu]
\end{align*}
where the second equality follows from Equation~\eqref{eq:YwTindepTU}.  Together, under Assumption~\ref{iv}~\ref{monotonicity}, we have:
\begin{align}
  & \E[Y_i(1, \bu) - Y_i(0, \bu) \mid \wT_i(1, \bu) =1, \wT_i(0, \bu)=0] \nonumber \\
  = \ & \frac{\E[Y_i \mid T_i = 1, \bU_i = \bu] - \E[Y_i \mid T_i = 0, \bU_i = \bu]}{\E[\wT_i  \mid T_i = 1, \bU_i = \bu] - \E[\wT_i \mid T_i = 1, \bU_i = \bu]}. \label{cond_late}
\end{align}

Finally, the LATE is identified as:
\begin{align*}
&\E[Y_i(1,\bU_i) - Y_i(0,\bU_i)\mid \wT_i(1,\bU_i) =1,  \wT_i(0,\bU_i)=0]\\
 = \ &\int_{\cU}\E[Y_i(1,\bU_i) - Y_i(0,\bU_i) \mid \wT_i(1, \bU_i) =1,  \wT_i(0, \bU_i) = 0, \bU_i] dF(\bU_i \mid \wT_i(1,\bU_i) =1, \wT_i(0,\bU_i)=0)\\
= \ & \int_{\cU} \frac{\E[Y_i \mid T_i = 1, \bU_i] - \E[Y_i \mid T_i = 0, \bU_i ]}{\E[\wT_i  \mid T_i = 1, \bU_i ] - \E[\wT_i \mid T_i = 1, \bU_i ]} dF(\bU_i \mid \wT_i(1,\bU_i) > \wT_i(0,\bU_i))\\
= \ & \int_{\cU} \frac{\E[Y_i \mid T_i = 1, \bU_i ] - \E[Y_i \mid T_i = 0, \bU_i ]}{\E[\wT_i  \mid T_i = 1, \bU_i ] - \E[\Tilde{T}_i \mid T_i = 1, \bU_i ]} \cdot \frac{\P(\wT_i(1,\bU_i) =1, \wT_i(0,\bU_i) = 0 \mid \bU_i )}{\P(\wT_i(1,\bU_i) =1, \wT_i(0,\bU_i)=0)} dF(\bU_i)\\
= \ & \int_{\cU} \frac{\E[Y_i \mid T_i = 1, \bU_i ] - \E[Y_i \mid T_i = 0, \bU_i ]}{\E[\wT_i  \mid T_i = 1, \bU_i ] - \E[\wT_i \mid T_i = 1, \bU_i ]} \cdot \frac{\E[\wT_i  \mid T_i = 1, \bU_i ] - \E[\wT_i \mid T_i = 1, \bU_i ]}{\P(\wT_i(1,\bU_i) =1, \wT_i(0,\bU_i)=0)} dF(\bU_i)\\
= \ & \frac{\int_{\cU} \E[Y_i \mid T_i = 1, \bU_i ] - \E[Y_i \mid T_i = 0, \bU_i ] dF(\bU_i)}{\int_{\cU} \E[\wT_i  \mid T_i = 1, \bU_i] - \E[\wT_i \mid T_i = 0, \bU_i] dF(\bU_i)}\\
= \ & 
\frac{\ \int_{\mathcal{R}} \E[Y_i \mid T_i = 1, 
    \boldf(\bR_i)] - \E[Y_i \mid T_i = 0, 
    \boldf(\bR_i)] dF(\bR_i)}{\int_{\mathcal{R}} \E[\wT_i  \mid T_i = 1, \boldf(\bR_i)] - \E[\wT_i \mid T_i = 0, \boldf(\bR_i)] dF(\bR_i)}
\end{align*}
where the second equality follows from Equation~\eqref{cond_late}, the third equality is due to Bayes' theorem, and the last equality (as well as the existence of deconfounder) follows from the same argument used to establish Theorem~\ref{iden_det}. \qed

\section{Results of Additional Simulation Studies} \label{app:sim}

\begin{table}[!h]
\centering \spacingset{1}
\begin{tabular}{l.....}
  \toprule
  & & & \multicolumn{2}{c}{95\% confidence interval} & \multicolumn{1}{c}{Runtime} \\
& \multicolumn{1}{c}{Bias} & \multicolumn{1}{c}{RMSE} & \multicolumn{1}{c}{coverage} & \multicolumn{1}{l}{avg. length} & \multicolumn{1}{c}{(seconds)} \\
\midrule
  \textbf{Weak confounding w/ separability} \\
Proposed estimator (new) & -0.33 & 1.06 & 0.94 & 3.47 & 42.1 \\ 
Proposed estimator (reuse) & -0.26  & 0.98 & 0.92 & 2.95 & 54.1 \\
Difference-in-Means & 3.61 & 3.61 & 0 & 4.72 &  0.0 \\ 
Outcome model with BERT & 1.17 & 1.00 & 0.12 & 0.53 & 296 \\ 
DML with BERT & 0.58 & 4.21 & 0.93 & 2.10 & 327\\ 
\midrule
\textbf{Moderate confounding w/ separability} \\
Proposed estimator (new) & -1.07 & 2.72 & 0.95 & 9.00 & 43.6 \\ 
Proposed estimator (reuse) & -1.05 & 2.36 & 0.93 & 6.85 & 62.9 \\
Difference-in-Means  & 7.95 & 7.95 & 0 & 9.50 & 0.0 \\ 
Outcome model with BERT & 3.44 & 2.27 & 0.05 & 0.92 & 673 \\ 
DML with BERT & 2.09 & 18.3 & 0.92 & 5.14 & 720 \\ 
\midrule 
\textbf{Strong confounding w/ separability} \\
Proposed estimator (new) & -14.6 & 36.9 & 0.88 & 113 & 55.3 \\ 
Proposed estimator (reuse) & -15.1 & 36.0 & 0.92 & 96.3 & 74.3 \\
Difference-in-Means  & 86.0 & 86.0 & 0 & 95.7 & 0.0 \\ 
Outcome model with BERT & 112 & 114 & 0 & 13.2 & 2731\\ 
DML with BERT & 208 & 917 & 0.26 & 382 & 2756 \\ 
\midrule
\textbf{Weak confounding w/o separability} \\
Proposed estimator (new) & 3.23 & 3.27 & 0 & 1.45 & 35.9 \\ 
Proposed estimator (reuse) & 2.87 & 2.89 & 0 & 1.34 & 41.1 \\
Difference-in-Means  & 2.20 & 2.20 & 0.03 & 4.03 & 0.0 \\
Outcome model with BERT & 2.55 & 2.70 & 0.01 & 1.02 & 320\\ 
DML with BERT &  6.18 & 16.7 & 0.05 & 4.69 & 348  \\ 
\midrule
  \textbf{Moderate confounding w/o separability} \\
Proposed estimator (new) & 6.70 & 6.76 & 0 & 2.89 & 42.5\\ 
Proposed estimator (reuse) & 5.90 & 5.93 & 0 & 2.66 & 44.9 \\
Difference-in-Means  & 4.39 & 4.40 & 0 & 8.04 & 0.0 \\
Outcome model with BERT & 7.56 & 7.66  & 0  & 1.85 & 624\\ 
DML with BERT & 12.1 & 30.5 & 0.06 & 10.1 & 656 \\ 
\midrule 
\textbf{Strong confounding w/o separability} \\
Proposed estimator (new) & 76.3 & 76.7 & 0 & 29.8 &  53.0 \\ 
Proposed estimator (reuse) & 66.8 & 67.1 & 0 & 27.7 & 56.4 \\
Difference-in-Means  & 44.0 & 44.0 & 0 & 80.3 & 0.0  \\ 
Outcome model with BERT & 116 & 117 & 0 & 13.2 & 2689  \\ 
DML with BERT & 207 & 814 & 0.26 & 425 & 2716  \\ 
\bottomrule
\end{tabular}
\caption{Simulation Results with 200 Monte Carlo trials}
\label{sim_result_wint}
\end{table}


\begin{table}[h]
\centering \spacingset{1}
\begin{tabular}{l.....}
  \toprule
  & & & \multicolumn{2}{c}{95\% confidence interval} & \multicolumn{1}{c}{Average} \\
& \multicolumn{1}{c}{Bias} & \multicolumn{1}{c}{RMSE} & \multicolumn{1}{c}{coverage} & \multicolumn{1}{l}{avg. length} & \multicolumn{1}{c}{time (sec.)} \\
\midrule
  \textbf{Weak confounding w/ separability} \\
Proposed estimator (new) & -0.31 & 1.09 & 0.93 & 3.55 & 43.0 \\ 
Proposed estimator (reuse) & -0.21 & 0.96 & 0.93 & 2.90 & 55.7 \\
  \midrule
  \textbf{Moderate confounding w/ separability} \\
Proposed estimator (new) & -0.99 & 2.77 & 0.92 & 8.83 & 45.7  \\ 
Proposed estimator (reuse) & -1.00 & 2.67 & 0.90 & 6.99 & 61.7 \\
\midrule
  \textbf{Strong confounding w/ separability} \\
Proposed estimator (new) & -14.3  & 36.1 & 0.89 & 108 & 57.7  \\ 
Proposed estimator (reuse) & -16.0 & 38.1 & 0.90 & 98.8 & 76.3 \\
\bottomrule
\end{tabular}
\caption{Simulation Results based on 1000 Monte Carlo trials}
\label{sim_result_1000}
\end{table}

\newpage
\section{Additional Empirical Application: Hong Kong Experiment} \label{app:hk}

To further validate the proposed GPI methodology, we apply it to the Hong Kong experiment conducted by \citet{fong_causal_2023}. This experiment examines the extent to which U.S. commitments to Hong Kong influence public perceptions of U.S. government support for Hong Kong protesters. To investigate this question, the authors carried out two experiments—one in December 2019 ($N = 1{,}983$) and another in October 2020 ($N = 2{,}072$). For each experiment, they first generated 555,660 unique candidate texts by randomly varying several text features: descriptions of commitments (commitments the United States made to Hong Kong), bravery (the bravery displayed by the protesters), mistreatment (China’s mistreatment of its own citizens), flags (whether protesters were shown waving American flags), threat (the security threat China poses to the United States), economy (information about Hong Kong’s political system and economy), and violations (how China’s actions violate its treaty with the United Kingdom).

To mitigate confounding bias arising from these text features, the authors created roughly 15 variants of texts for each feature, randomly concatenated two or three variants to form a complete text, and then randomly assigned the resulting texts to participants. Participants read a text and then rated, on a 0–100 scale, how strongly they agreed with the view that the U.S. government should support Hong Kong protesters. Because textual features are randomized, the authors regressed participants’ responses on the seven text features using ordinary least squares (OLS).

We estimate the average treatment effect separately for each experimental wave. Following the empirical application in Section~\ref{sec:empirical}, we use the GPI methodology based on the text-reuse approach with Llama 3-8B. After extracting the internal representation, we apply the proposed estimation procedure described in Section~\ref{sec:methodology}, using five-fold cross-fitting. We adopt a larger fold size than in Section~\ref{sec:empirical} to ensure that the neural network is trained on a sufficiently large number of samples. For comparison—consistent with our simulation studies in Section~\ref{sec:simulation} and the empirical application in Section~\ref{sec:empirical}—we also implement two existing BERT-based methods: the outcome-model approach with BERT \citep{pryzant_causal_2021} and DML with BERT \citep{gui_causal_2023}. Finally, we replicate the original OLS analysis. In this application, the experimental design should effectively mitigate confounding bias, so OLS serves as a reasonable approximation to the ground truth.

\begin{table}[t]
    \centering \spacingset{1}
    \begin{tabular}{l.c..}
    \toprule
    Methods & \multicolumn{1}{c}{ATE Estimates} & \multicolumn{1}{c}{95\% Confidence Interval} & \multicolumn{1}{c}{IOSS} & \multicolumn{1}{c}{Runtime  (sec.)}   \\
    \midrule
    \textbf{Wave 1: December 2019} & & & & \\
    OLS (original) & 5.231 & [1.814, \hspace{.04in} 8.648] & $-$ & 0.0 \\
    GPI  (reuse) & 6.175 & [2.784, \hspace{.04in} 9.566]  & 0.04 & 35.9 \\
    Outcome model with BERT & 26.591 & [25.482, 27.701]  & \multirow{2}{*}{\makebox[0pt][l]{\hspace*{-.08in}0.28}} & 11890.6 \\
    DML with BERT & 24.361 & [20.163,  28.560]  &  & 11892.9 \\
    \midrule
    \textbf{Wave 2: October 2020} & & & & \\
    OLS (original) & 2.680 & [0.269, \hspace{.04in} 5.091]  & $-$ & 0.0 \\
    GPI  (reuse) & 2.043 & [-0.790, \hspace{.04in} 4.877]  & 0.04 & 27.9 \\
    Outcome model with BERT & 1.676 & [1.319, 2.033]  & \multirow{2}{*}{\makebox[0pt][l]{\hspace*{-.08in}0.07}} & 9940.0 \\
    DML with BERT &  2.808 & [$-$0.519,  6.136]  & & 9942.7 \\
    \bottomrule
    \end{tabular}
    \caption{The Estimated Average Treatment Effect (ATE) for the Hong Kong Experiment.}
    \label{experiment2}
\end{table}

Table~\ref{experiment2} presents the estimated ATEs for both experimental waves. The GPI estimates are consistent with the original OLS estimates that controls the textual features directly. We find that IOSS for the GPI method is close to 0 (0.04 for both waves), indicating that the deconfounder extracted by GPI is disentangled from the treatment feature. In contrast, the BERT-based methods yield significantly larger ATE estimates for Wave~1 that significantly diverge from the OLS estimates. This unually large estiamte corresponds to a large value of IOSS for Wave~1, which is 0.28 (the IQSS score for Wave~2 is smaller, equaling 0.07). Moreover, the GPI method is computationally efficient, with runtimes significantly faster than those of the BERT-based methods.

\end{document}